\documentclass[a4paper,fleqn,usenatbib,useAMS]{mnras}

\usepackage{graphicx}	
\usepackage{amsmath}	
\usepackage{amssymb}	
\usepackage{multicol}
\usepackage{bm}		
\usepackage{pdflscape}	
\usepackage[T1]{fontenc}
\usepackage{aecompl}

\usepackage{multirow}

\newcommand{\kms}{\,km\,s$^{-1}$} 
\newcommand{\kmsM}{\,km\,s$^{-1}$\,Mpc$^{-1}$} 

\usepackage[T1]{fontenc}
\usepackage{ae,aecompl}

\usepackage{mathptmx}

\title[Role of major mergers in stellar mass growth of BCGs]{The close pair fraction of BCGs since $z=0.5$: major mergers dominate recent BCG stellar mass growth}

\author[D. N.~Groenewald et al.]{Dani\`{e}l N.~Groenewald$^{1,2}$\thanks{E-mail: dgroenewald@saao.ac.za},
Rosalind E.~Skelton$^{1}$,
David G.~Gilbank$^1$ \newauthor
and S. Ilani~Loubser$^2$
\\
$^{1}$South African Astronomical Observatory, Observatory Road, 7925, Cape Town, South Africa\\
$^{2}$North-West University, Potchefstroom, 2520, South Africa
}

\date{Accepted XXX. Received YYY; in original form ZZZ}

\pubyear{2017}

\begin{document}
\label{firstpage}
\pagerange{\pageref{firstpage}--\pageref{lastpage}}
\maketitle

\begin{abstract}
Using the redMaPPer cluster catalogue based on the Sloan Digital Sky Survey (SDSS) photometry, we investigate the importance of major mergers in the stellar mass build-up of brightest cluster galaxies (BCGs) between $0.08 \leq z \leq 0.50$. We use the SDSS spectroscopy, supplemented with spectroscopic observations from the Southern African Large Telescope at higher redshifts, to identify which BCGs and nearby companions are potential major merger candidates. We use the pair fraction as a proxy for the merger fraction in order to determine how much stellar mass growth the BCGs have experienced due to major mergers. We observe a weak trend of the BCG pair fraction increasing with decreasing redshift, suggesting that major mergers may become more important towards the present day. Major mergers are found to contribute, on average, $24 \pm 14 $ $(29 \pm 17)$ per cent towards the stellar mass of a present day BCG since $z=0.32$ (0.45), assuming that half of the companion's stellar mass is accreted onto the BCG. Furthermore, using our merger results in conjunction with predictions from two recent semi-analytical models along with observational measurements from the literature, we find that major mergers have sufficient stellar material to account for the stellar mass growth of the intracluster light between $z=0.3$ and $z=0$. 
\end{abstract}



\section{Introduction}
\label{sec:intro}

Brightest cluster galaxies (BCGs) are the most massive and luminous galaxies in the Universe. These galaxies are typically found close to, or at, the centres of their host clusters \citep{Jones1984, Jones1999, Beers1986, Rhee1991}. In the hierarchical formation scenario BCGs are thought to form through the accretion of smaller galaxies \citep[e.g.][]{Ostriker1977, Richstone1983}. These massive galaxies are also known to have properties that are very different from that of other early-type galaxies (ETGs). In comparison to other ETGs of the same mass, BCGs are found to be larger \citep[][and references therein]{Bernardi2007} and have extended light profiles \citep{Matthews1964, Tonry1987, Schombert1988, Gonzalez2000, Gonzalez2003}. These unique properties have been attributed to their special location at the cluster centres \citep{Hausman1978}. These properties may also suggest that the formation of BCGs are different from that of other ETGs \citep[e.g.][]{Burke2000, Stott2008, Stott2010}. 

Due to the privileged positions of BCGs at the centres of their host clusters, these massive galaxies are expected to experience multiple mergers during their lifetime, making them ideal probes with which to study galaxy formation. Observational evidence, for example tidal tails and distorted isophotes, indicates that some BCGs are experiencing mergers \citep[e.g.][]{Bernardi2007, Lauer2007, 2007, McIntosh2008, Liu2009, Liu2015, Rasmussen2010, Brough2011}, but it is not clear to what extent their stellar mass is being built up by mergers and at what rate. 

There is currently no consensus in the literature regarding the stellar mass evolution of BCGs between $z=1$ and the present day. Observational studies have found that these massive galaxies changed their mass by factors that range between one \citep[equivalent to no growth; e.g.][]{Aragon-Salamanca1998, Whiley2008, Collins2009, Stott2010, Oliva-Altamirano2014} and $1.4 \pm 0.2$ \citep{Lin2013} or $1.8 \pm 0.3$ \citep{Lidman2012} since $z \lesssim 1$.

Predictions from numerical simulations and semi-analytical models (SAMs) suggest that BCGs form in two `phases'. They suggest that star formation dominates the BCG's mass growth at $z\geq2$, while multiple dry mergers of smaller galaxies dominate the mass assembly at $z\leq1$ \citep[e.g.][]{DeLucia2007, Naab2009, Laporte2013}. The simulations, however, differ in the stellar mass growth they predict for BCGs. For example, \cite{DeLucia2007} find that BCGs have changed their mass by a factor of 4 from $z = 1 - 0$, mainly through minor\footnote{Typically defined in the literature as mergers with stellar mass ratios of $>$ 1:4 $-$ 1:20 \citep[e.g.][]{Edwards2012, Burke2013, Burke2015}.} mergers. \cite{Laporte2013} on the other hand predict a BCG mass growth factor closer to 2 over the same redshift range, however this model finds that both major\footnote{Typically defined as mergers with stellar mass ratios of 1:1 to 1:4 \citep[e.g.][]{Jogee2009,Lopez-Sanjuan2012,Robotham2014}.} and minor mergers contribute significantly towards the mass growth of these massive galaxies.

The discrepancies between the BCGs' observed stellar mass growth and that predicted by simulations are quite apparent. This may be because mergers do not contribute significantly to the stellar mass growth of the BCGs if a significant fraction of the merging mass ends up in the cluster's intracluster light (ICL). The ICL can be described as stars that are not gravitationally bound to any single galaxy, but rather to the cluster potential. Although the origin of the ICL is not known, it is thought that tidal stripping of satellite galaxies and merger events (involving BCGs) contribute towards the stellar mass growth of the ICL. The ICL is often found to be more concentrated around the central galaxy in a cluster \citep{Mihos2005, Rudick2011}. This in turn implies that the formation and evolution of BCGs and the ICL are connected to one another. 

The recent simulation of \cite{Laporte2013} predicts that 30 per cent of a companion galaxy's mass will be distributed into the ICL during a merger with the BCG \citep[see also][]{Conroy2007, Puchwein2010}. This brings the model's predicted BCG mass growth over $0 < z < 1$ into better agreement with the observed stellar mass growth estimates of \citet{Lidman2012, Lin2013, Burke2013} in the same redshift range. If we are to believe that some fraction of the merging mass ends up in the ICL, then this presents us with a scenario that the ICL is being built-up through galaxies that are interacting with the BCGs. \cite{DeMaio2015} however argues against this idea, stating that the ICL is being built-up by the stripping of satellite galaxies with luminosities $>0.2 \, L^{\star}$ \cite[also see][]{Contini2014}. Consequently, it is clear that the growth and stellar mass build-up of BCGs and the ICL are linked, however it is not clear to what extent merger events contribute towards this.

\medskip 
In this paper we investigate the importance of major mergers in the stellar mass build-up of BCGs between $0.08 \leq z \leq 0.50$. We select BCGs from a photometric cluster catalogue, which has been constructed from the Sloan Digital Sky Survey \citep[SDSS;][]{York2000}. We use spectroscopic information from the SDSS to identify which BCGs and nearby companions are potential major merger candidates. The close pair fraction is used as a proxy for the merger fraction to determine how much stellar mass growth the BCGs will experience due to mergers.

This paper is structured as follows. Section \ref{sec:sample_descr} contains details pertaining to the redMaPPer catalogue; in Section \ref{sec:method} the methods are described. The results on the pair fraction and merger-inferred stellar mass growth of the BCGs are shown and discussed in Section \ref{sec:results_disc}. Conclusions are drawn in Section \ref{sec:concl}. Throughout the paper we assume a flat $\Lambda$CDM cosmology with $\Omega_{M} = 1 - \Omega_{\Lambda} = 0.3$ and $H_{0} = 70$ \kmsM. Magnitudes are given in the AB system and the \cite{Chabrier2003} initial mass function is used. 

\section{Data}
\label{sec:sample_descr}

\subsection{Overview of the redMaPPer cluster catalogue}
\label{sec:redM_over}

Throughout this paper we use version 5.2 of the red-sequence Matched-filter Probabilistic Percolation cluster catalogue \citep[redMaPPer\footnote{\href{http://risa.stanford.edu/redmapper/}{http://risa.stanford.edu/redmapper/}};][]{Rykoff2014}. This catalogue covers roughly 10 000 $\mathrm{deg^{2}}$ of the sky and consists of more than 25\,000 galaxy clusters which span a redshift range of $0.08 \leq z \leq 0.55$. The clusters in this catalogue have been optically identified using photometry from the Eighth Data Release \citep[DR8;][]{Aihara2011} of the SDSS. Where possible, we supplement the catalogue with additional spectroscopic redshifts from the SDSS DR12 \citep{Alam2015}. A detailed description of the redMaPPer cluster catalogue (and construction) can be found in \cite{Rykoff2014}. Here, we summarize the most important features. 

Briefly, the redMaPPer cluster detection algorithm photometrically identifies clusters by searching for overdensities of red-sequence galaxies. This relies on a set of galaxies with spectroscopic redshifts that are used to construct a redshift dependent red-sequence model. The spectroscopic redshifts needed for the calibration of the red-sequence model are retrieved from the SDSS Main Galaxy Sample \citep[MGS;][]{Strauss2002}, Luminous Red Galaxy \citep[LRG;][]{Eisenstein2001}, and Baryon Oscillation Spectroscopic Survey \citep[BOSS;][]{Ahn2012}. This model is then used to photometrically group red galaxies (with luminosities $\geq 0.2 \, L^{\star}$) at similar redshifts into clusters, assuming a radial filter that corresponds to typical cluster sizes. The algorithm iteratively determines a photometric redshift for each cluster based on the calibrated red-sequence model, as measured from the photometrically identified candidate cluster members.

\subsubsection{Probability of being a cluster member}
\label{sec:clus_mem}

The candidate cluster members are each assigned to a redMaPPer cluster based on a membership probability ($P_{\rm{MEM}}$). $P_{\rm{MEM}}$ indicates the probability of a galaxy being a red-sequence galaxy that belongs to a specific cluster. This follows directly from the method used by the cluster detection algorithm to identify clusters, i.e. looking for overdensities of red-sequence galaxies, with the luminosity cut and radial filtering mentioned above. redMaPPer provides catalogues of (candidate) cluster members for each cluster, based on the galaxies' individual properties, including $P_{\rm{MEM}}$, and it is these catalogues which we use throughout our analysis.

\begin{figure*}
\centering
\includegraphics[width=15cm, height=13cm]{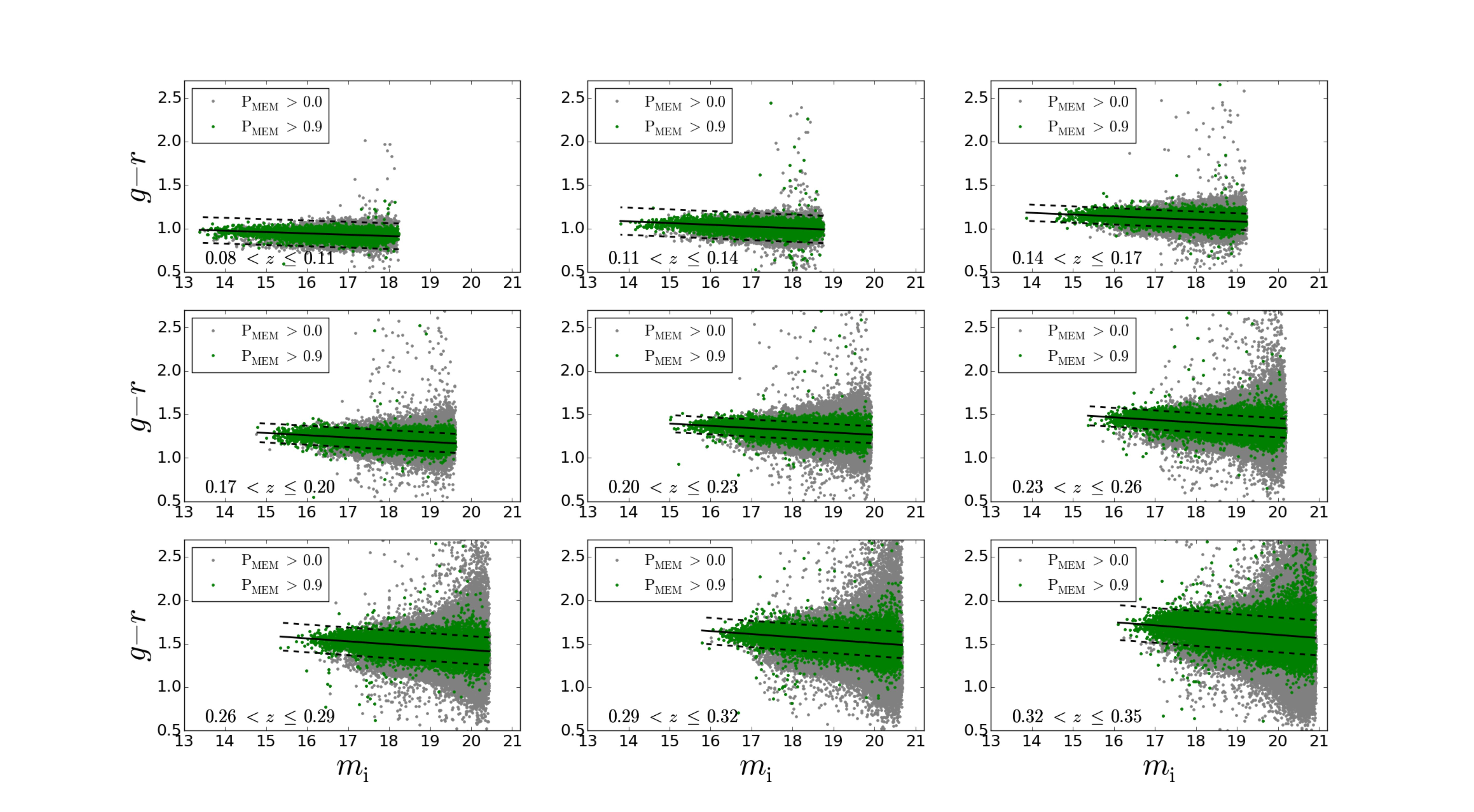}
\caption[]{The colour-magnitude diagram of the redMaPPer clusters in various redshift bins. All candidate cluster members and those with $P_{\rm{MEM}} > 0.9$ are shown in grey and green respectively. The solid and dashed lines respectively indicate the location of the red-sequence and the $1\sigma$ width. Only high probability cluster members ($P_{\rm{MEM}} > 0.9$) have been used to find the red-sequence, since these galaxies are more likely to belong to the red-sequence. The red-sequence forms a tighter sequence as $P_{\rm{MEM}}$ increases and becomes redder with redshift.}
	\label{fig:PR_rms_cuts_red_seq}
\end{figure*}

In Fig. \ref{fig:PR_rms_cuts_red_seq} we show the colour-magnitude diagrams ($g-r$ vs. $m_\mathrm{i}$) of the redMaPPer clusters out to $z=0.35$ (the redshift cut will be explained in the next section). All candidate cluster members and those with $P_{\rm{MEM}} > 0.9$ are shown in grey and green respectively. In each panel it is apparent that members with large $P_{\rm{MEM}}$ values form a tighter sequence, as expected, since these galaxies are more likely to be red-sequence galaxies. We will return to this idea in Section \ref{sec:fpair_results}.

For each cluster in the redMaPPer catalogue, the central galaxy identification algorithm of redMaPPer assigns five galaxies probabilities of being the central galaxy (CG) of the cluster ($P_{\rm{CEN}}$). These galaxies are ranked according to probability and are hereafter referred to as the CG candidates. The central probabilities are defined by using a luminosity filter, photometric redshift filter and a local galaxy density filter (discussed below). The product of these three filters produces the overall centering filter that is used to determine the central probabilities \citep[see equation 67 of][]{Rykoff2014}. The redshift filter that is used in this equation is slightly broader than the cluster red-sequence filter in order to allow galaxies with slight colour offsets from the red-sequence to be considered as CG candidates, as it is possible for the CGs to have experienced residual amounts of star formation and therefore have bluer photometric colours \citep[$\sim2$ per cent of the CGs in redMaPPer are blue;][]{Rykoff2014}. 

The galaxy with the highest $P_{\mathrm{CEN}}$ is not necessarily the brightest galaxy. This is because, apart from the luminosity, the local galaxy density around the CG candidate is also considered. Central galaxies are expected to be found in the highest density central regions of clusters. Consequently, the redMaPPer algorithm gives higher preference to galaxies in denser regions than those in less dense regions. A less luminous galaxy located in a denser environment than the most luminous galaxy may therefore have a higher $P_{\mathrm{CEN}}$.

The total magnitudes of the galaxies in the redMaPPer catalogue are given by the $i$-band \texttt{cModel\_Mag} \citep[$m_{\mathrm{i}}$;][]{Abazajian2004}, while the colours of the galaxies are determined using \texttt{modelMag} \citep{Abazajian2004} in the $u, g, r, i$ and $z$-bands. All magnitudes and colours have been corrected for Galactic extinction using the dust maps of \cite*{Schlegel1998}. 

\subsubsection{Richness of the redMaPPer clusters}
\label{sec:richn}

\begin{figure*}
\centering
\includegraphics[scale=0.6]{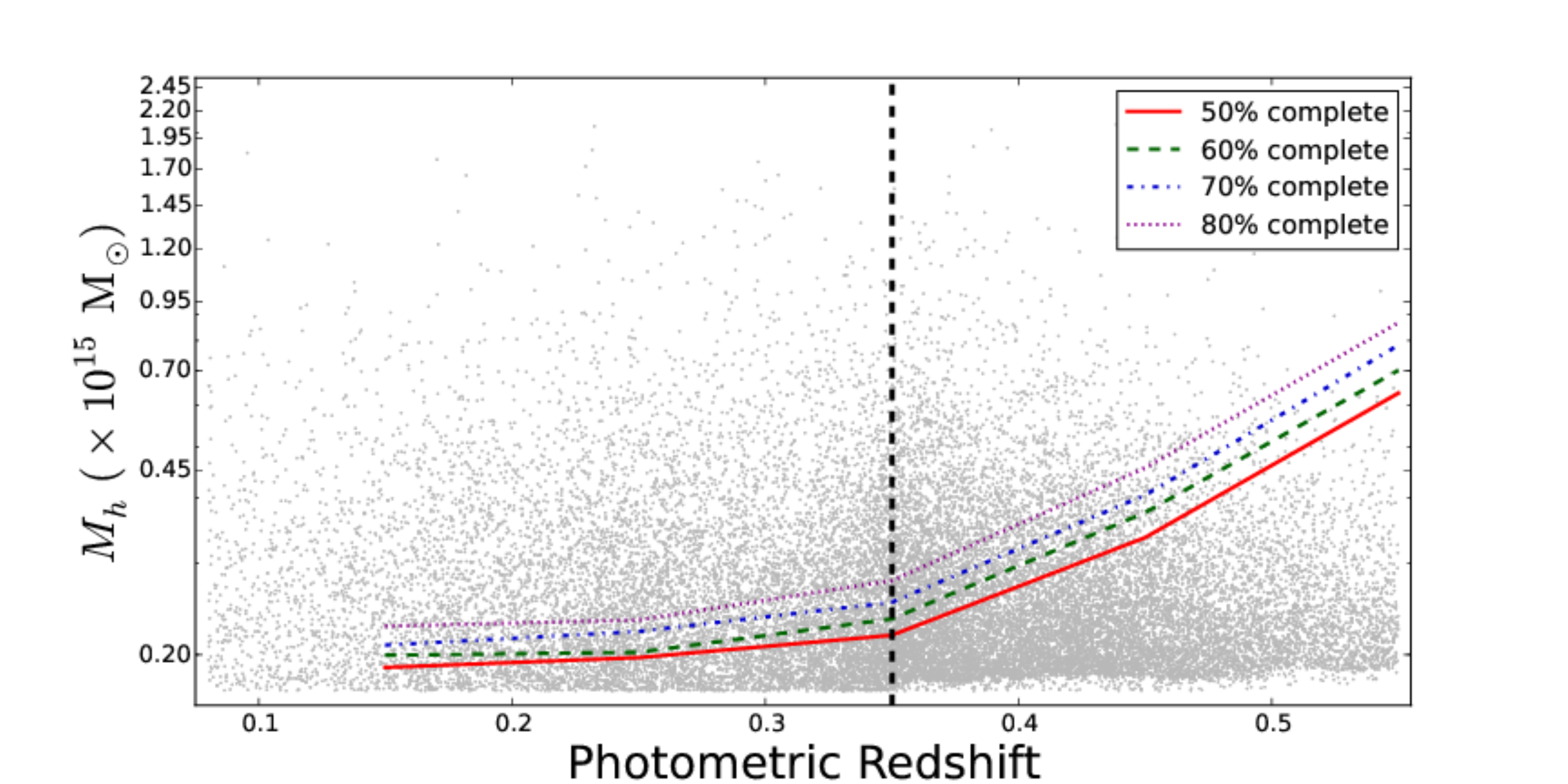}
\caption[]{Completeness of the redMaPPer catalogue as a function of $M_{h}$ and redshift. The grey points represent the $M_{h}$-distribution of the clusters in the catalogue (derived using the corrected richnesses). The various lines are constructed from the values in figure 22 of \cite{Rykoff2014}. The redMaPPer algorithm used five redshift bins to determine the completeness of the catalogue as a function of richness. The lines join the mean redshift value of each bin. At $z\leq0.35$ (vertical dashed line), the galaxy catalogue is volume limited and complete down to clusters with $M_{h} \sim 0.2 \times 10^{15} \, \mathrm{M_{\odot}}$. At $z > 0.35$, however, the algorithm is only able to detect the most massive clusters due to the magnitude limit of the survey and is therefore incomplete for low mass (low richness) clusters. See text for details.}
	\label{fig:redM_completeness_fig}
\end{figure*}

The catalogue contains clusters with richnesses\footnote{Richness is defined as the number of red-sequence galaxies within a cluster that are brighter than $0.2 \, L^{\star}$.} of $\lambda \geq 20 \, S(z)$, where $S(z)$\footnote{Given by equation 23 of \cite{Rykoff2014}.} is a correction factor that is used to account for the survey depth of the SDSS. We show the effect of the survey depth on the measured completeness of the redMaPPer clusters as a function of their halo masses $(M_{h})$ in Fig. \ref{fig:redM_completeness_fig}. The $M_{h}$ have been derived using the halo mass-richness relation of \cite{Rykoff2012}, assuming richness\footnote{We correct the redMaPPer richness estimates for the survey depth using $\lambda / S(z)$. We use these in the remainder of the paper and refer to them as the corrected richnesses.} is a proxy for $M_{h}$. The various lines in Fig. \ref{fig:redM_completeness_fig} are constructed from the values in figure 22 of \cite{Rykoff2014}. The redMaPPer algorithm used five redshift bins to determine the completeness of the catalogue as a function of richness. The lines in this figure join the mean redshift value of each bin. At $z\leq0.35$, the galaxy catalogue is volume limited and the survey depth is brighter than the fiducial luminosity cut of $0.2 \, L^{\star}$ (therefore $S(z)=1$). At these redshifts the cluster catalogue is 50 per cent complete down to clusters with $M_{h} \gtrsim 0.2 \times 10^{15} \, \mathrm{M_{\odot}}$. At $z>0.35$, however, the magnitude limit of the survey causes only the most massive clusters to be detected. This causes the richness detection threshold to increase with redshift. 

In this work, we restrict our analysis to only consider clusters in the volume limited sample at $z\leq0.35$. We further only consider redMaPPer clusters with halo masses above the 50 per cent $M_{h}$ completeness limit. This is done to maximize the number of clusters in the evolutionary sequence (as discussed in Section \ref{sec:evol_seq_clusters}). 

\subsubsection{BCGs - Identification and stellar masses}
\label{sec:BCG_selection}

In our work we select the first CG candidate (the galaxy with the highest $P_{\rm{CEN}}$) in each cluster as the BCG candidate.

The redMaPPer catalogue does not supply any stellar mass $(M_{\star})$ estimates for the galaxies. We determine stellar mass estimates using version 4.1 of the {\tt kcorrect} code \citep*{Blanton2007} which uses templates based on the \cite{Bruzual2003} models. Briefly, spectral energy distribution fitting is used to derive the $M_{\star}$ of the galaxies by fitting their observed \texttt{modelMag} magnitudes (in the $u, g, r, i $ and $z$-bands) against a range of spectral templates from \cite{Bruzual2003} assuming a \cite{Chabrier2003} initial mass function. The best-fitting stellar template is determined through $\chi^{2}$-minimization, whereafter a mass-to-light ratio is determined from this template and used to convert the luminosity of the galaxy to stellar mass. 

\subsection{The SALT sample}

We extend the BCG pair fraction analysis out to moderately high redshifts ($0.4 \leq z \leq 0.5$) using the Gaussian Mixture Brightest Cluster Galaxy catalogue \citep[GMBCG\footnote{\href{http://home.fnal.gov/~jghao/gmbcg_sdss_catalog.html}{http://home.fnal.gov/$\tt{\sim}$jghao/gmbcg$\_$sdss$\_$catalog.html}};][]{Hao2010} with follow-up spectroscopy from the Southern African Large Telescope \citep[SALT;][]{Buckley2006, ODonoghue2006}. We require a sample with a high spectroscopic completeness in order to identify pairs that are likely to merge. The spectroscopic completeness of the SDSS drops significantly around $z=0.4$, so we have supplemented the SDSS data with additional redshifts from SALT. 

The GMBCG cluster catalogue is a photometrically-identified cluster sample similar to redMaPPer and was used in the early stages of this work. When redMaPPer became available, we switched to using this due to several advantages it offered, such as explicit BCG probabilities, candidate member galaxy probabilities, etc. However, spectroscopic follow-up was already well underway for the GMBCG sample. In this work, we treat them as two independent cluster samples, in the same way as comparing our redMaPPer work with those of others in the literature, and will show that our results are compatible. 

We refer the reader to \cite{Hao2010} for details on the GMBCG catalogue. Briefly, the catalogue spans a redshift range of $0.1 < z < 0.55$ and contains 55 000 clusters, optically identified from the SDSS DR7 \citep{Abazajian2009} as overdensities of red-sequence galaxies. The positions of the BCGs and the photometric redshifts of their host clusters are provided in the catalogue. Here we are interested in the 4814 BCGs that are found in the redshift range $0.4 \leq z \leq 0.5$. 

We use the method outlined in Section \ref{sec:pair_selection} to select the close pairs that are included in the SALT sample. Briefly, all companions within a physical separation distance of 50 kpc and 1.5 magnitudes (in the $i$-band) of the BCGs were retrieved from the DR7 database. The spectroscopic redshifts (if available) of these galaxies were retrieved from the SDSS DR12 \citep{Alam2015}. We further restricted the SALT sample to only include pairs where either the BCG or companion (in each pair) had a SDSS spectroscopic redshift $(z_{\mathrm{spec}})$. SALT observations were used to determine the remaining galaxy's $z_{\mathrm{spec}}$ to determine whether the galaxies are potential merger candidates or not. Only 16 of the close pairs satisfied these criteria. Twelve pairs were successfully observed over two semesters spanning 2013 November - 2014 April (proposal ID: 2013-2-RSA-008, PI: Groenewald) and  2014 May - October (proposal ID: 2014-1-RSA\_OTH-009, PI: Groenewald).

\subsubsection{Observations and reductions}

Using the Robert Stobie Spectrograph \citep[RSS;][]{Burgh2003, Kobulnicky2003} on SALT we obtained longslit spectroscopy for the close pairs. A slit with a width of $2\arcsec$ was centred on the BCG in each pair and aligned in such a way that both the BCG and companion were observed in a single observation. This allowed us to determine the relative velocities between these two galaxies using the same wavelength calibration. The RSS observations used the PG900 grating which covers the main optical features that we are interested in over $4500 - 7500$ \AA. Each close pair in the SALT sample was observed for a total of 106 minutes, split over two observation blocks (each observation block consisted of $2 \times 20$ minute exposures).

Basic data reductions, i.e. gain and cross-talk correction as well as bias subtraction were performed as part of the automated reduction pipeline of SALT \citep{Crawford2010}. We performed cosmic ray rejection on the science images by using the \texttt{LACosmic} package \citep{2001}. Wavelength calibrations were then performed with standard \texttt{IRAF} \citep{Tody1986, Tody1993} tasks. We determined the spectroscopic redshift of each galaxy in the SALT sample by fitting the observed SALT spectrum against the `Early-type' SDSS galaxy spectral template (hereafter reference spectrum). The rest wavelengths of the Calcium II $H$ and $K$ absorption lines in the reference spectrum were shifted to match those in the galaxy's observed SALT spectrum. 

In Table \ref{table:SALT_sample} we present a summary of the galaxies in the SALT sample along with their SDSS spectroscopic redshifts (if available). For each close pair we indicate the velocity difference as derived from the SALT spectroscopy. The BCG and companion in each pair were observed in the same slit and share the same wavelength calibration. Thus the velocity differences, which are our ultimate goal, are more robust than if we had used a combination of SALT and SDSS redshifts for each pair.  

\begin{table*}
\caption[]{Summary of the close galaxy pairs in the SALT sample. The pair ID of each pair (taken from the GMBCG catalogue) is given in Column 1 while the galaxies' coordinates are given in Columns 3 and 4. The SDSS spectroscopic redshifts of these galaxies, if available, are given in Column 5. The velocity difference of each close pair, determined from SALT spectroscopy, is given in Column 6.}
 \label{table:SALT_sample}
 \begin{tabular}{lccccc} 
 		\hline
		
 		\multicolumn{1}{l}{Pair ID} &
 		\multicolumn{1}{c}{} &
 		\multicolumn{1}{c}{RA (J2000)} &
 		\multicolumn{1}{c}{DEC (J2000)} &	
 		\multicolumn{1}{c}{$z_{\mathrm{SDSS}}$} &
 		\multicolumn{1}{c}{$\Delta v_{\mathrm{SALT}}$} \\
		
 		\multicolumn{1}{c}{} &
 		\multicolumn{1}{c}{} &
 		\multicolumn{1}{c}{(deg)} &
 		\multicolumn{1}{c}{(deg)} &
 		\multicolumn{1}{c}{} &
 		\multicolumn{1}{c}{(\kms)} \\
        
 		\hline
		
 		\multirow{2}{*}{337} & BCG & 190.404850 & $-$0.668280 & $0.4587 \pm 0.0001$ & \multirow{2}{*}{$21 \pm 85$} \\  
 		& Companion & 190.406416 & $-$0.667039 & --- & \\ 

 		\multirow{2}{*}{5501} & BCG & 128.317079 & 0.108010 & --- & \multirow{2}{*}{$1017 \pm 85$} \\ 
 		& Companion & 128.316370 & 0.108480 & $0.4748 \pm 0.0001$ &  \\ 

 		\multirow{2}{*}{5919} & BCG & 153.259460 & 0.812003 & $0.4063 \pm 0.0001$ & \multirow{2}{*}{$4517 \pm 85$} \\ 
 		& Companion & 153.259030 & 0.809660  & --- &  \\ 

 		\multirow{2}{*}{22105} & BCG & 27.261019 & $-$0.639814 & $0.3584 \pm 0.0001$ & \multirow{2}{*}{$685 \pm 42$} \\ 
 		& Companion & 27.261605 & $-$0.639844 & --- & \\ 

 		\multirow{2}{*}{22258} & BCG & 346.949880 & 0.948270 & $0.3686 \pm 0.0001$ & \multirow{2}{*}{$307 \pm 85$} \\ 
 		& Companion & 346.949133 & 0.947189 & --- &  \\ 

 		\multirow{2}{*}{23941} & BCG & 186.754036 & 0.765942 & --- & \multirow{2}{*}{$2612 \pm 42$} \\ 
 		& Companion & 186.753290 & 0.767780 & $0.5153 \pm 0.0002$ &  \\ 

 		\multirow{2}{*}{24097} & BCG & 148.316850 & 1.272180 & $0.3648 \pm 0.0001$ & \multirow{2}{*}{$198 \pm 42$}  \\ 
 		& Companion & 148.316048 & 1.272079 & --- &  \\ 

 		\multirow{2}{*}{431} & BCG & 206.032182 & 1.948250 & --- & \multirow{2}{*}{$39 \pm 42$}\\ 
 		& Companion & 206.031830 & 1.948720 & $0.5432 \pm 0.0001$ &   \\ 
	
 		\multirow{2}{*}{5680} & BCG & 223.098640 & 0.949810 & $0.4830 \pm 0.0001$ & \multirow{2}{*}{$40 \pm 85$} \\ 
 		& Companion & 223.099826 & 0.950909 & --- &   \\ 

 		\multirow{2}{*}{5905} & BCG & 210.260650 & 0.275680 & --- & \multirow{2}{*}{$61 \pm 42$} \\ 
 		& Companion & 210.258767 & 0.275656 & $0.4757 \pm 0.0001$ & \\ 

 		\multirow{2}{*}{24726} & BCG & 218.763280 & 3.109750 & $0.3829 \pm 0.0001$ & \multirow{2}{*}{$152 \pm 85$} \\ 
 		& Companion & 218.763793 & 3.111125 & --- &  \\ 

 		\multirow{2}{*}{52685} & BCG & 357.091165 & 0.741055 & --- & \multirow{2}{*}{$106 \pm 85$} \\ 
 		& Companion & 357.088710 & 0.741300 & $0.4113 \pm 0.0001$ & \\ 
 		
 		\hline
 \end{tabular}
 \end{table*}

\section{Method - The pair fraction and merger-inferred stellar mass growth of the BCGs}
\label{sec:method}

\subsection{Constructing an evolutionary cluster sequence}
\label{sec:evol_seq_clusters}

The stellar masses of BCGs are known to correlate with the halo masses of their host clusters, with more massive halos hosting more massive BCGs \citep[e.g.][]{Edge1991, Burke2000, Brough2008, Stott2008, Stott2010, Stott2012, Whiley2008, Collins2009, Hansen2009, Lidman2012}. Consequently, it is important to take the halo mass growth of the clusters into account when the stellar mass growth of BCGs is investigated, to ensure that that BCGs at high redshifts are compared to their likely descendants at lower redshifts. This idea has already been explored by other works in the literature \citep{Lidman2012, Lin2013, Oliva-Altamirano2014, Zhang2016}. In our work we use an approach similar to that implemented by \cite{Lidman2012} to construct an evolutionary cluster sequence using evolving $M_{h}$ limits. We assume that the BCGs in this evolutionary sequence are progenitors/descendants of one another to derive the merger-inferred mass growth of the BCGs.

Briefly, we construct an evolutionary cluster sequence by identifying the low redshift descendants of the redMaPPer clusters in our highest redshift bin $(z=0.35)$. This is done by evolving the $M_{h}$ of these clusters forward in time using the mean mass accretion rates (MMAR) from the \cite*{Fakhouri2010} model (see their equation 2). Through these MMARs we are able to determine what the corresponding halo masses of these high redshift clusters will be at later times, allowing us to construct an evolving $M_{h}$ limit as a function of redshift. We divide the clusters into four equal sized redshift bins with a width of 0.067. The bin width is chosen to be larger than the typical uncertainty on the photometric redshift of the clusters ($\sim0.02$), which reduces the chances that clusters will be scattered in and out of adjacent redshift bins. Secondly, it is small enough to ensure we have multiple bins, each with a robust number of clusters, with which to study the BCGs' merger-inferred stellar mass growth. 

\begin{figure*}
\centering
\includegraphics[scale=0.6]{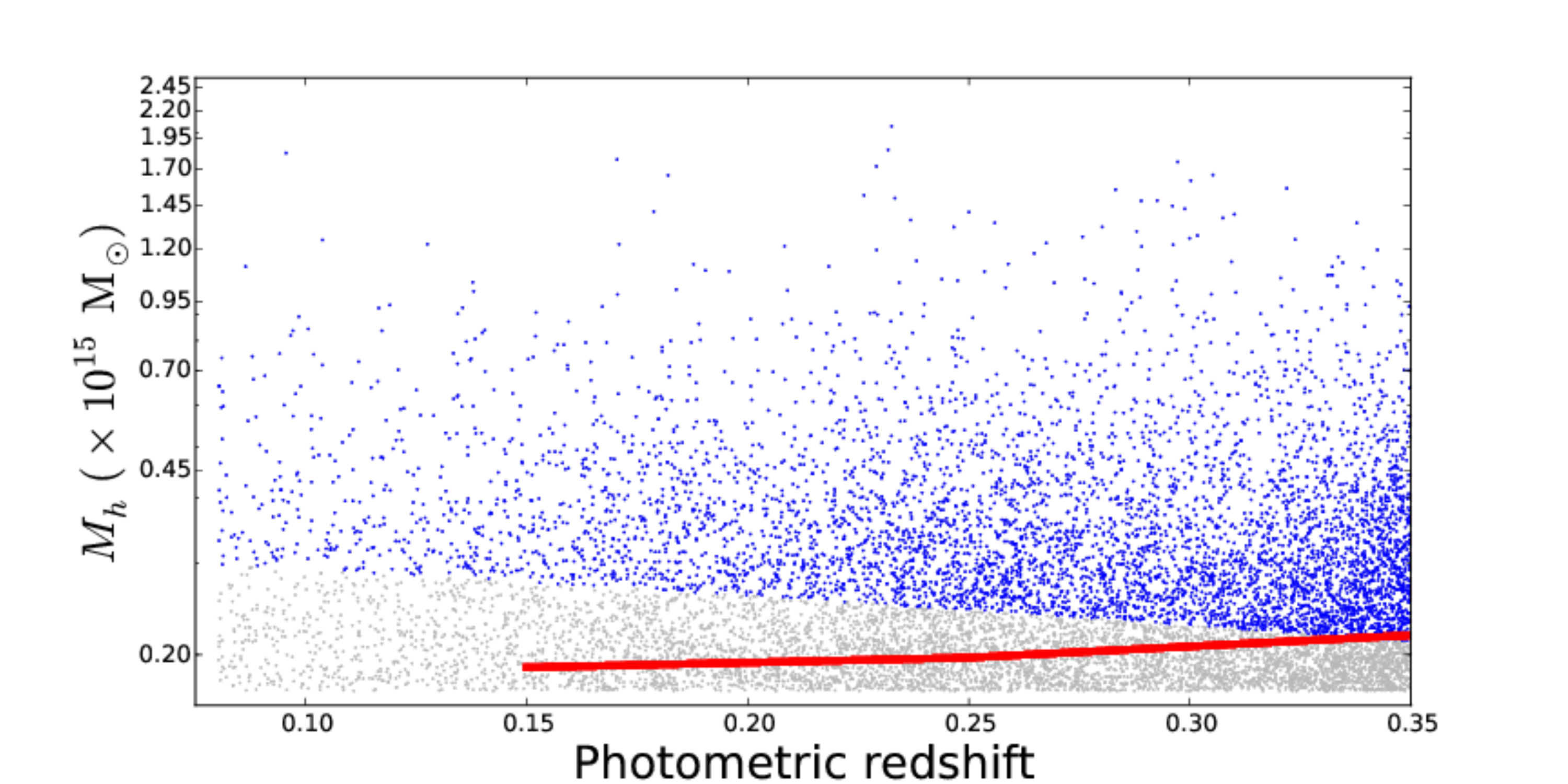}
\caption[]{The evolutionary cluster sequence of the redMaPPer clusters at $z=0.35$ that have been constructed using the MMARs of \cite{Fakhouri2010}. The $M_{h}$-distribution of the redMaPPer clusters is represented with the grey points. The clusters that form part of the sequence are indicated in blue, while the red line represents the 50 per cent $M_{h}$ completeness limit. See text for details.}
	\label{fig:Mhalo_match_no_upper_50}
\end{figure*}

We find a total of 5432 clusters that form part of the evolutionary sequence. These clusters (along with the evolving $M_{h}$ limit) are shown in Fig. \ref{fig:Mhalo_match_no_upper_50}. In Table \ref{table:f_pair} we present a summary of the number of clusters in each redshift bin along with their $M_{h}$ ranges. In Section \ref{sec:fpair_results} we test how the pair fraction of the BCGs is influenced when the halo mass growth of the clusters is not taken into account.

\subsection{The close pair selection}
\label{sec:pair_selection}

We begin by only considering galaxies that are brighter than 21.5 magnitudes in the $i$-band. This is the magnitude where the SDSS is 95 per cent complete for galaxies. This limit was determined by comparing the galaxy number counts in a region on the celestial equator from the SDSS to the Stripe 82 survey \citep{Adelman-McCarthy2007}, which is $\sim2$~mag deeper.

From the \textit{redMaPPer catalogue} we construct a close galaxy pair sample of 1336 pairs by searching for all photometric galaxies within a physical separation distance $(r_{\mathrm{sep}})$ of $7 \leq r_{\mathrm{sep}} \leq 50$ kpc from the BCGs. Galaxies within this search radius are referred to as companions. The lower limit is imposed since this is the minimum separation distance down to which the SDSS can resolve individual galaxies over our redshift range.

A close pair sample that has only been constructed using photometric information will inevitably suffer from contamination due to line-of-sight projections. Although the contamination in clusters is higher than in the field (including contamination from cluster members themselves), the same techniques can be used to obtain the close pair fraction. In this work we use spectroscopy to correct for this contamination by determining whether the photometrically identified companion galaxies are bound to their host BCGs. In order for close pairs to be considered potential merger candidates, we require the galaxies to have a velocity difference of $\Delta v \leq 300$\kms\ \citep[see e.g.][]{Burbidge1975, Ellison2013, Kitzbichler2008}. We have chosen this cut, which is somewhat stricter than many other observational works \citep[e.g.][]{Lin2004a, Lin2008, Lopez-Sanjuan2012, Robotham2014}, in order to use the results of \citet{Kitzbichler2008} to estimate merging timescales for each pair. These field merger timescales can be applied to close pairs in clusters \citep[see section 4.2 of][for motivation]{Lidman2013}.

\begin{figure*}
\centering
\includegraphics[width=15cm, height=12cm]{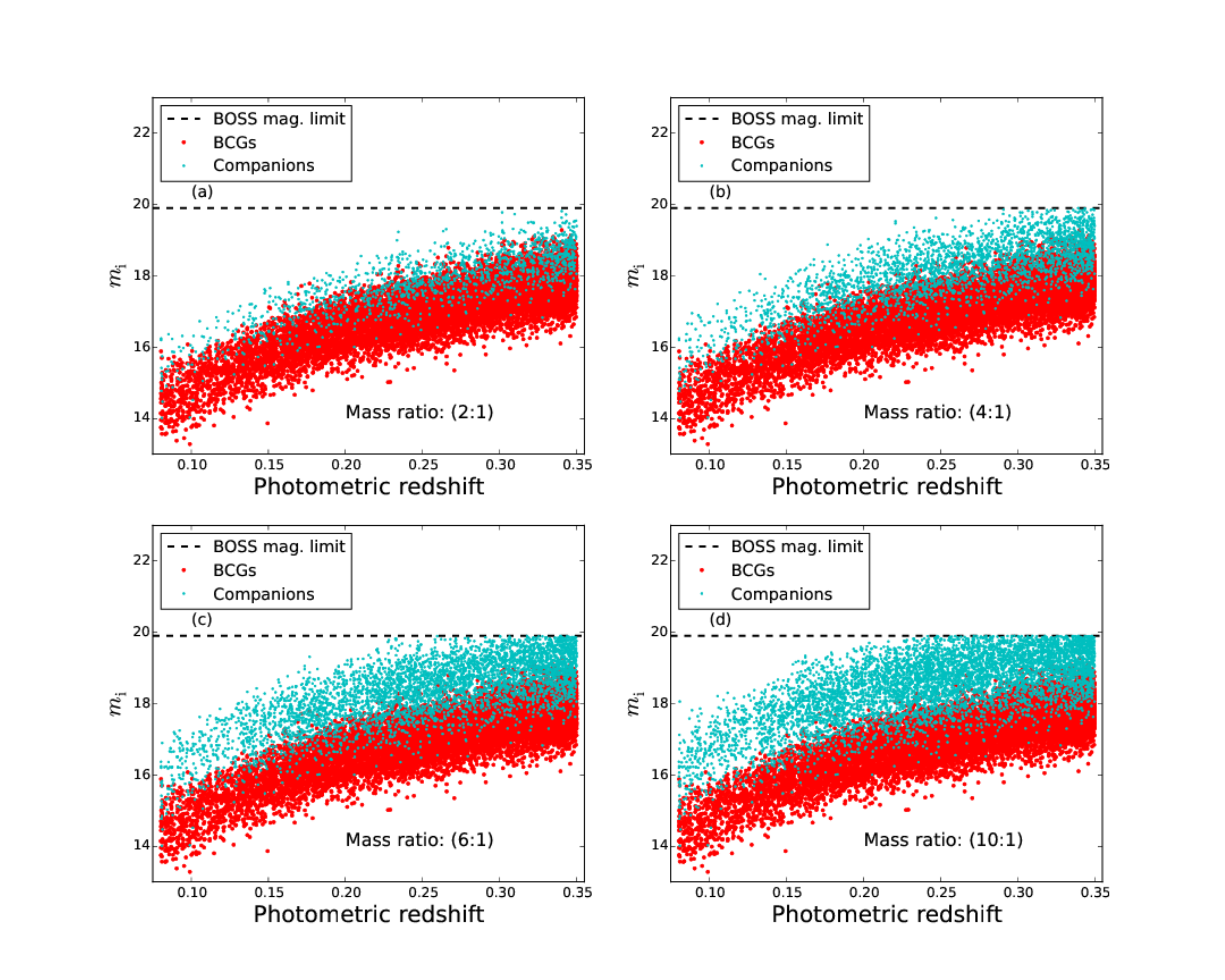}
\caption[]{To ensure we have a sample with a high spectroscopic completeness with which the BCG pair fraction can be determined, we require all the galaxies in our close pair sample to be brighter than the BOSS magnitude limit of $m_{\mathrm{i}}$ = 19.9 magnitudes (dashed line). We further restrict our sample to only include close pairs with stellar mass ratios of 1:1$-$1:4. This was done to maximize both the redshift range and number of close pairs in our sample. See text for details.}
	\label{fig:mag_lim_21_5}
\end{figure*}

\begin{figure*}
\centering
\includegraphics[scale=0.3]{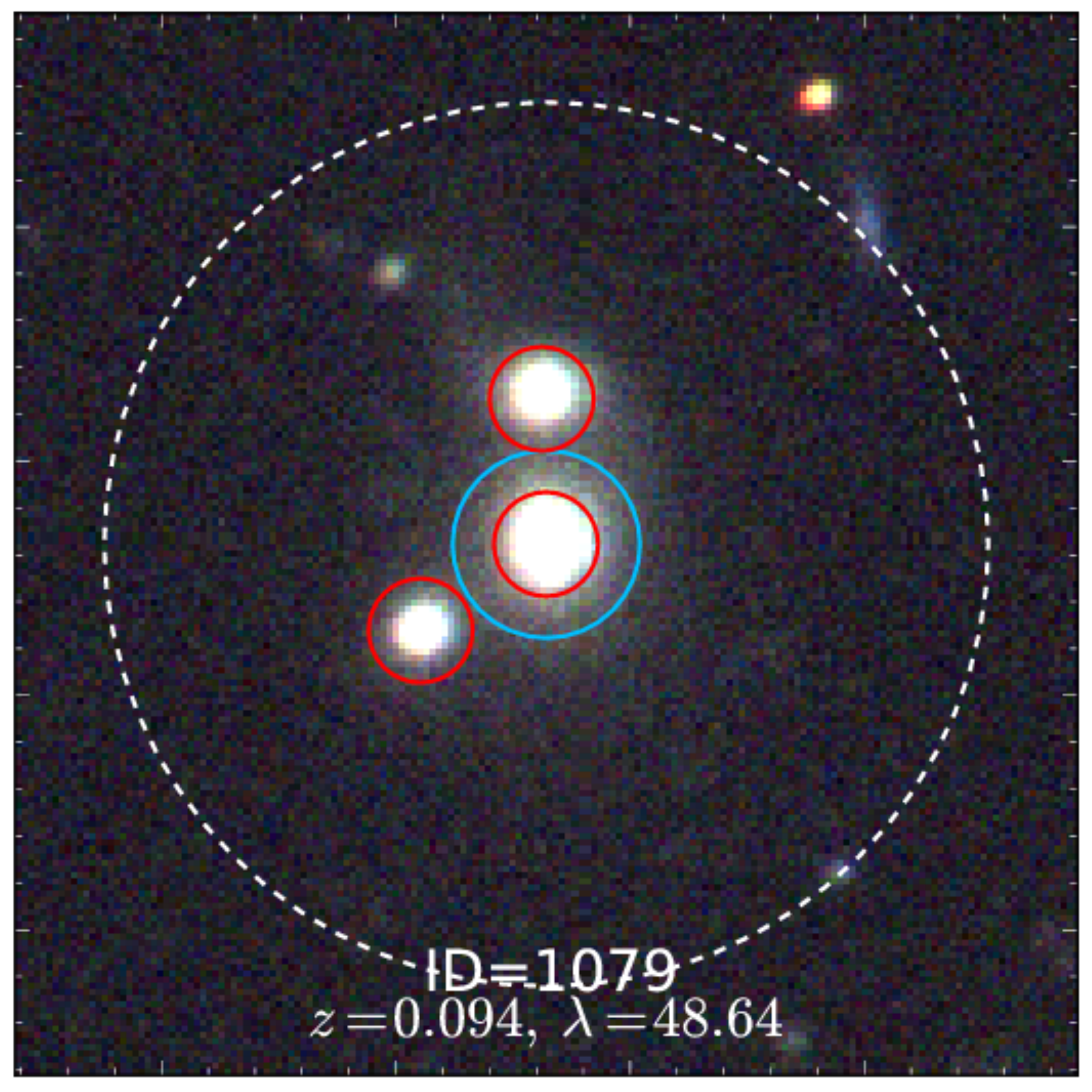}
\includegraphics[scale=0.3]{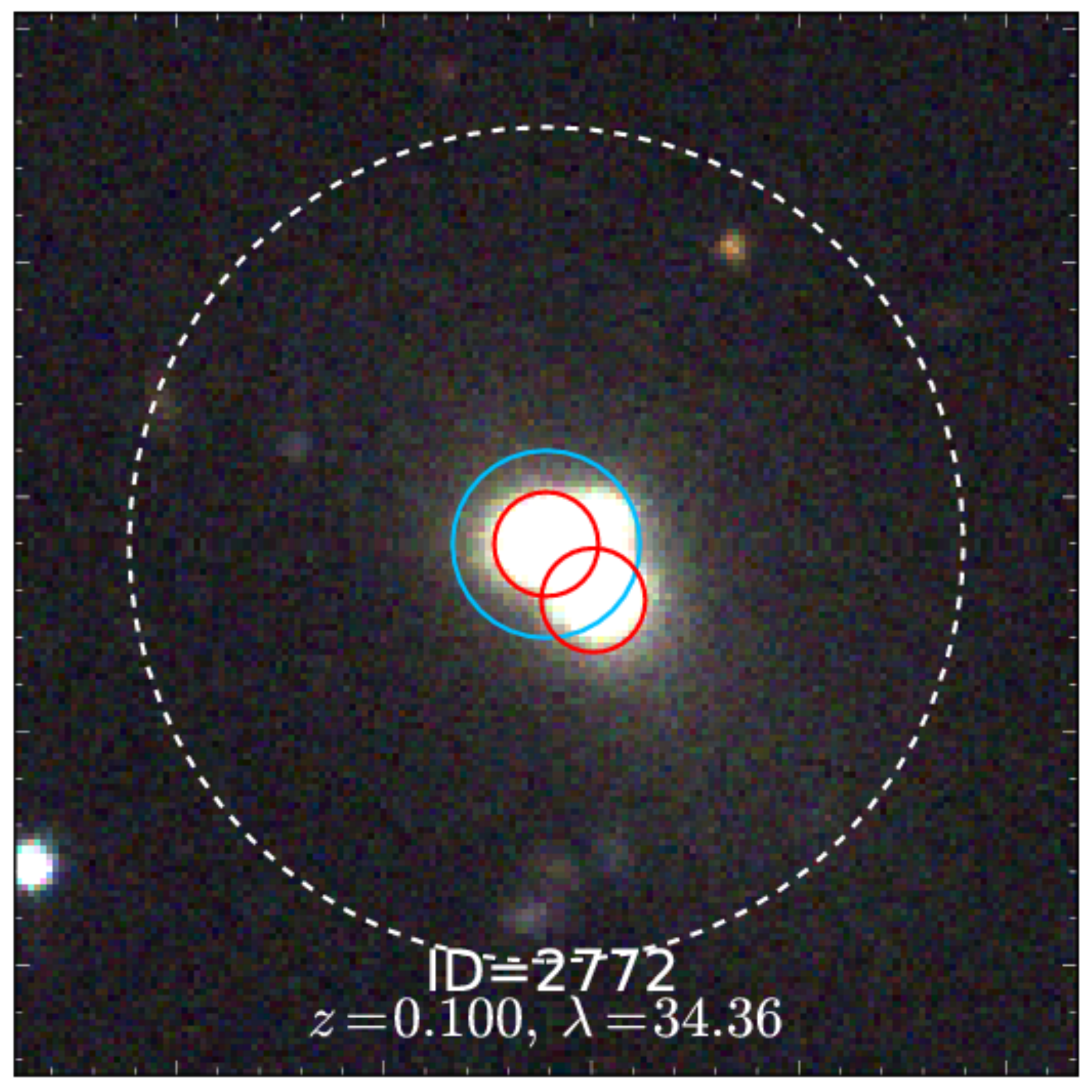}
\includegraphics[scale=0.3]{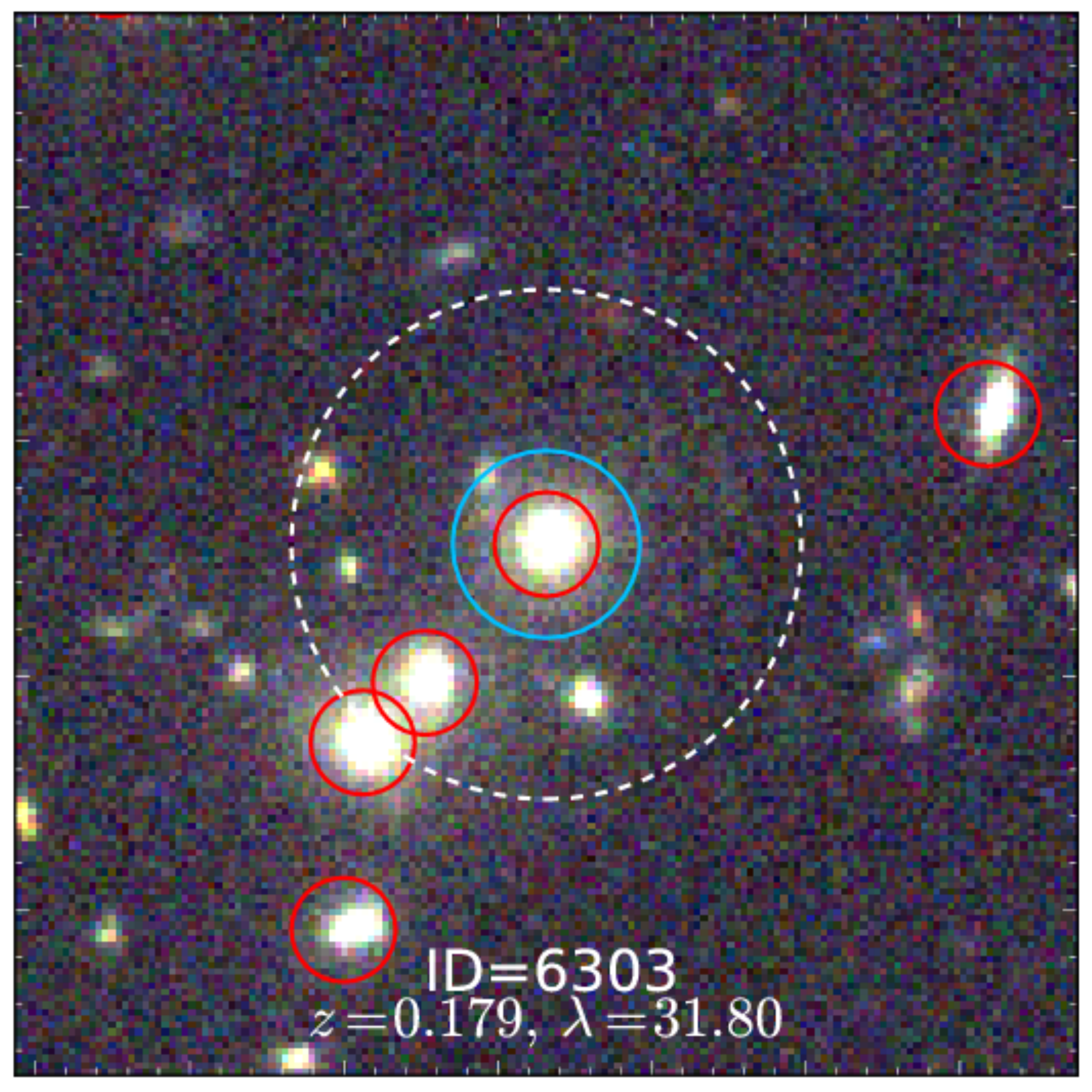}\\
\includegraphics[scale=0.3]{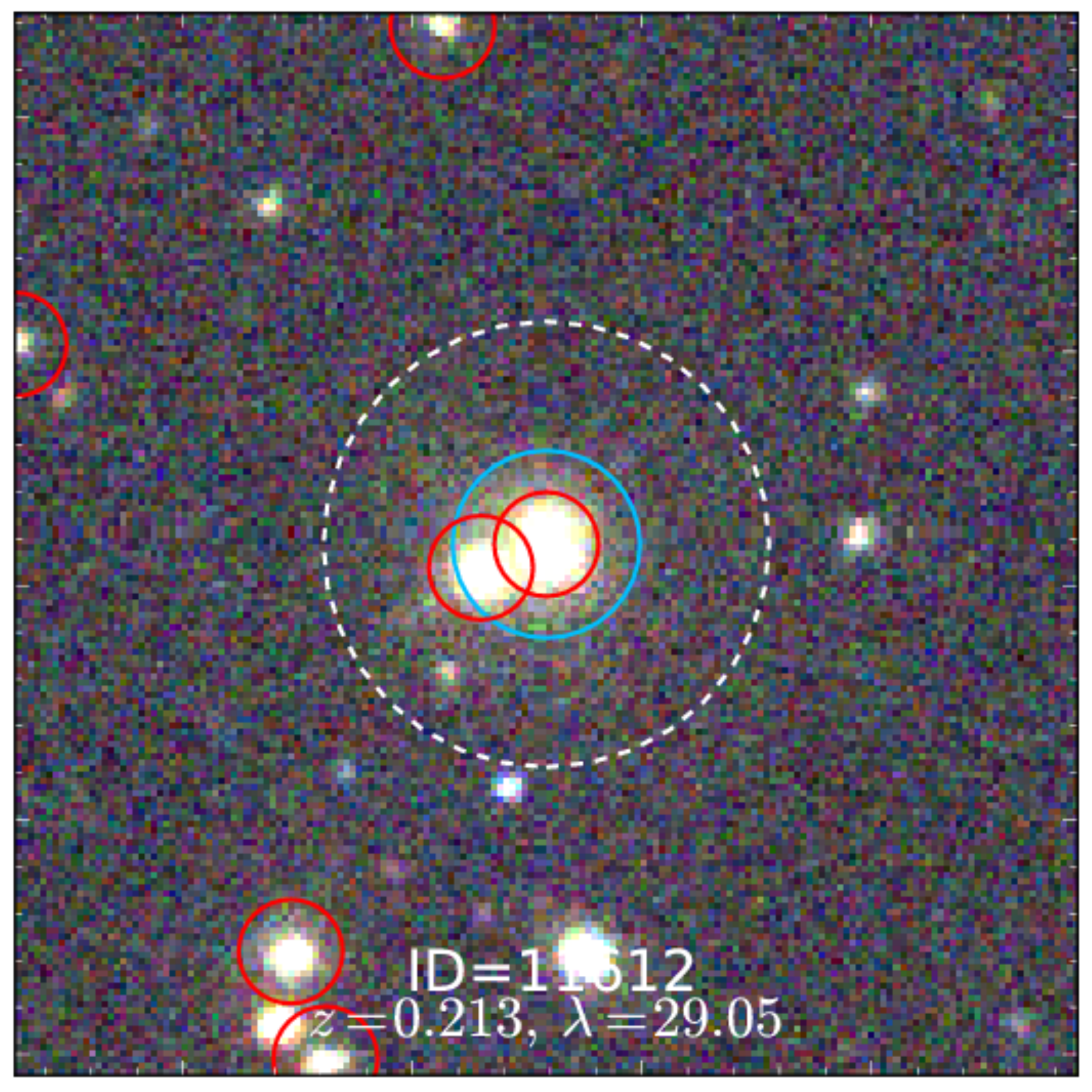}
\includegraphics[scale=0.3]{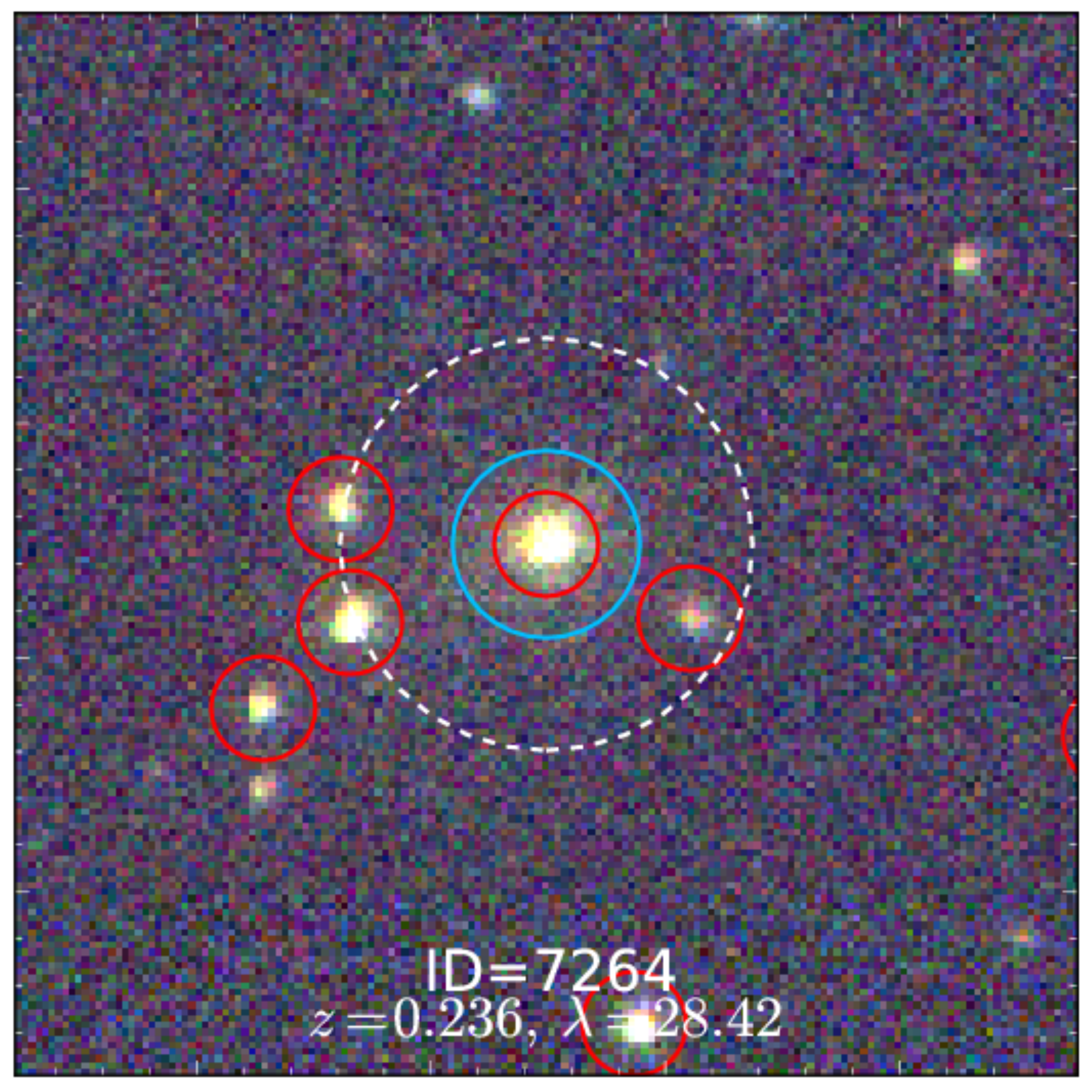}
\includegraphics[scale=0.3]{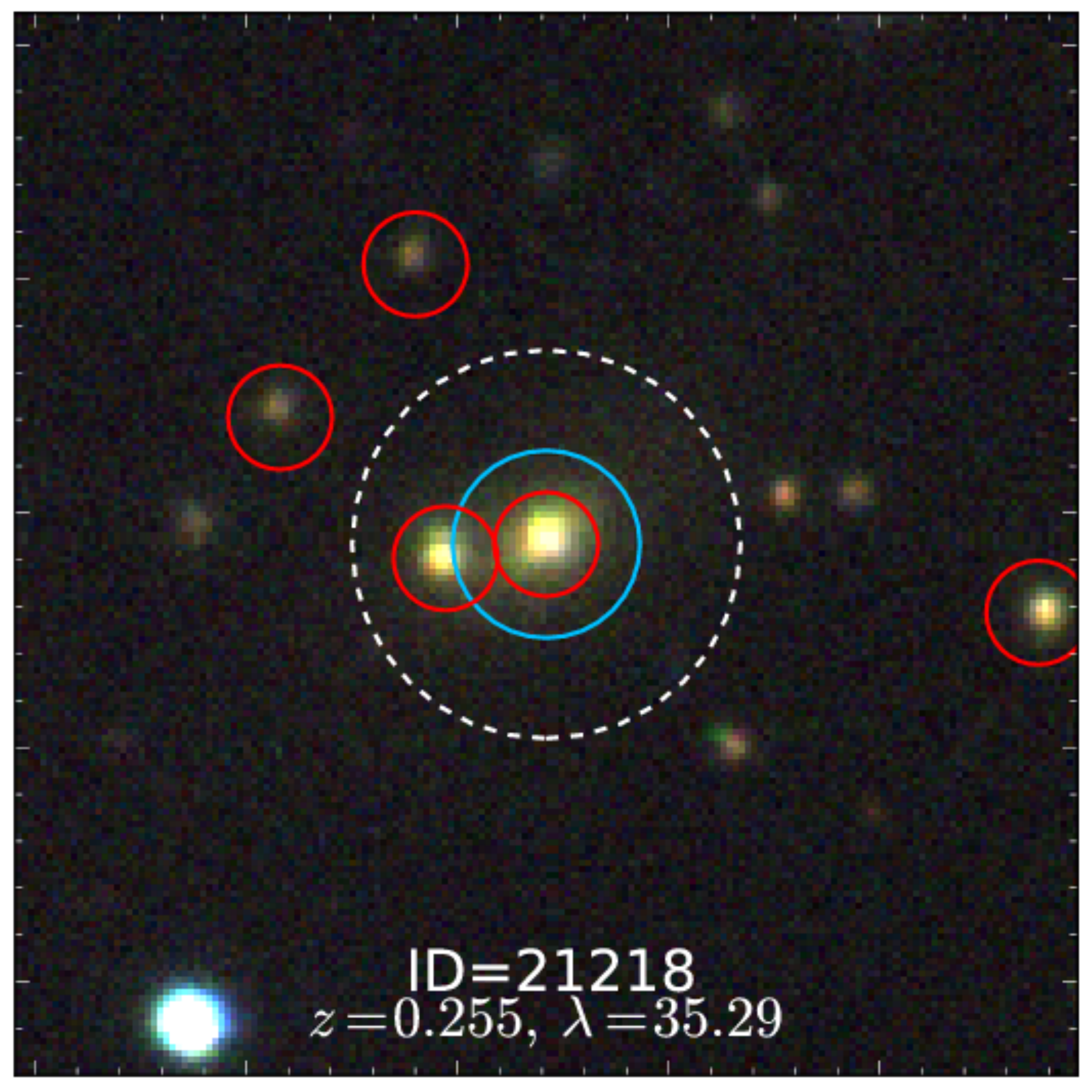}\\
\includegraphics[scale=0.3]{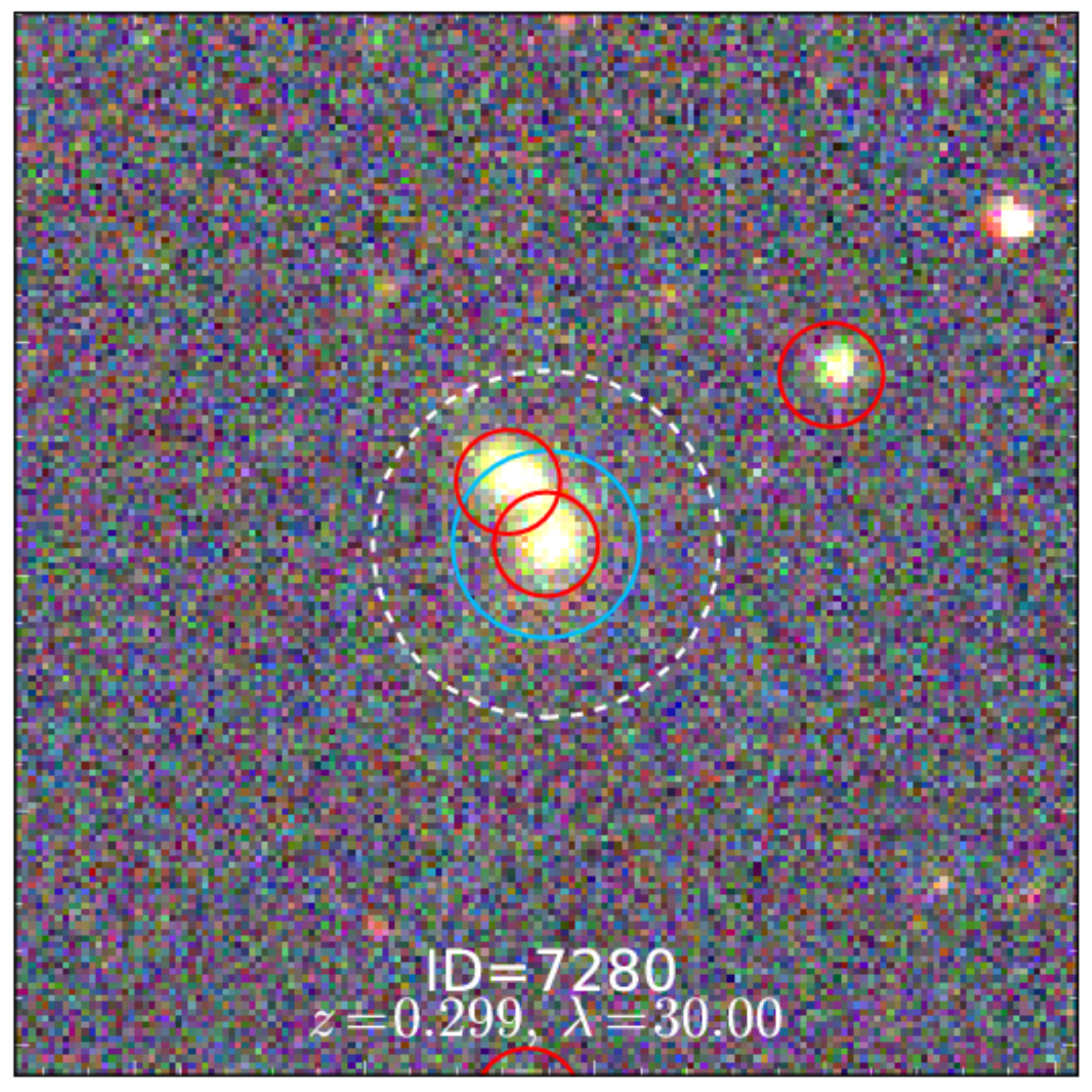}
\includegraphics[scale=0.3]{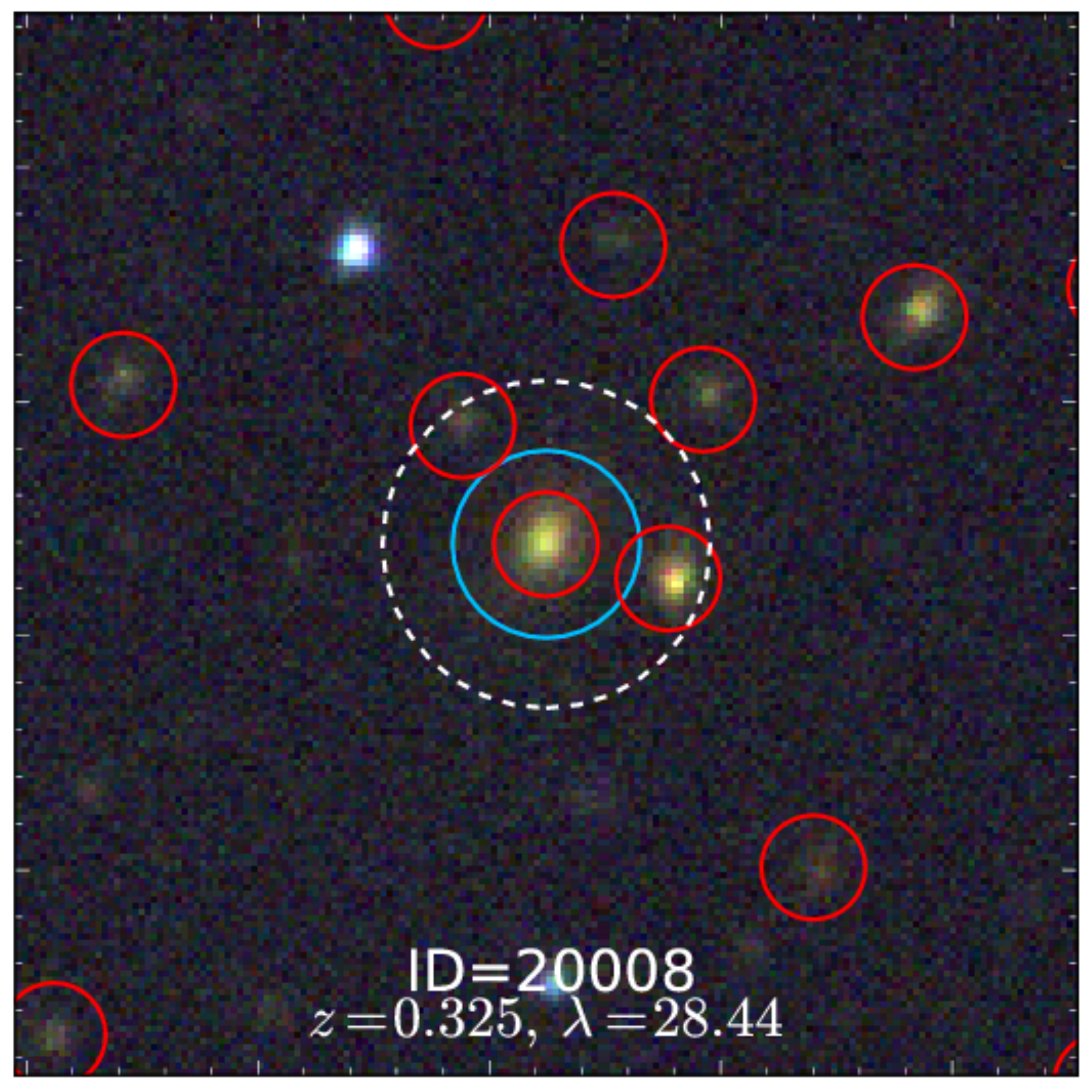}
\caption{For illustrative purposes, we show the SDSS cutouts of eight potential merger candidates (two per redshift bin) from our sample. Each cutout is centred on the BCG (circled in blue) and is $68\arcsec$ on a side, which roughly corresponds to 100 kpc at $z=0.08$. The objects circled in red are the redMaPPer candidate cluster members that are brighter than the luminosity cut of $0.2 \, L^{\star}$ (imposed during the construction of the catalogue). The dashed white line indicates the 50 kpc physical search radius (at the cluster's photometric redshift). For each cutout, we show the redMaPPer ID and richness along with the cluster's photometric redshift.}
\label{fig:SDSS_cutouts}
\end{figure*}

We use a compilation of the SDSS spectroscopic galaxy samples of which the BOSS survey supplies the deepest spectroscopy, up to a magnitude limit of 19.9 magnitudes in the $i$-band. We therefore further restrict our close pair sample to only include pairs with galaxies brighter than the BOSS magnitude limit. Our final sample consists of 1016 photometric pairs, of which 320 ($\sim31$ per cent) have spectroscopy. We correct for the spectroscopic incompleteness of our sample when the BCG pair fraction is measured (Section \ref{sec:fpair}). To ensure that we are relatively complete for companions to the BOSS magnitude limit, we only consider neighbouring galaxies with a stellar mass ratio within 1:4 of the BCG (see panel b of Fig. \ref{fig:mag_lim_21_5}). This maximizes the redshift range over which we are complete. We define all potential merger candidates with stellar mass ratios between 1:1$-$1:4 as major mergers. For illustrative purposes, we show the SDSS cutouts of eight major merger candidates from our close pair sample in Fig. \ref{fig:SDSS_cutouts}.

\subsection{Defining the pair fraction}
\label{sec:fpair}

We ultimately want to determine the fraction of BCGs that are in bound pairs, i.e. BCGs that will merge with their companions by $z=0$ (hereafter loosely referred to as `bound companions'). This is defined as follows: 
\begin{equation}
f_{\mathrm{pair}} \, = \, \frac{\mathrm{N_{BCGs,BC}}}{\mathrm{N_{BCGs}}} 
\label{eq:bound_merg_frac}
\end{equation}
where $\mathrm{N_{BCGs}}$ is the total number of BCGs that form part of the evolutionary sequence (as derived in Section~\ref{sec:evol_seq_clusters}) and $\mathrm{N_{BCGs,BC}}$ represents the number of BCGs with bound major merger companions.
We can expand Eq. \ref{eq:bound_merg_frac} to the following:
\begin{equation}
f_{\mathrm{pair}} \, = \, \frac{\mathrm{N_{BCGs,C}}}{\mathrm{N_{BCGs}}} \times \frac{\mathrm{N_{BCGs,BC}}}{\mathrm{N_{BCGs, C}}}
\label{eq:merg_frac}
\end{equation} 
where the first term is the fraction of BCGs with companions and the second term is the fraction of these companions that are bound (contamination correction). $\mathrm{N_{BCGs,C}}$ is the number of BCGs that have one or more companion(s) as described in Section \ref{sec:pair_selection}. In the case where only photometry is available, the second term (contamination correction) in Eq. \ref{eq:merg_frac} may be set to a constant value \citep[e.g. 0.5 in ][]{Edwards2012}. 

We obtain a measurement for the BCG pair fraction in two cases. We start with the simple case, where only photometry is used. This is used in studies which do not have spectroscopy available \citep[such as][]{Edwards2012}. Thereafter we consider the case where there is spectroscopic information. 

Where spectroscopic information is available for some of the sample, we have more information from which to determine whether companions are bound to the BCGs. Rather than assuming a constant value for the contamination correction in Eq. \ref{eq:merg_frac}, we derive a more detailed correction by grouping galaxies into bins of colours, magnitude and separation distance. In the $i$-th bin the pair fraction in redshift bin $j$ is
\begin{equation}
(f_{\mathrm{pair}})_{j} \, = \, \frac{\sum_{i} \, C_{i}^{-1} \, \mathrm{N_{BCGs,SC}}_{i}}{\sum_{i} \, \mathrm{N_{BCGs}}_{i}}
\label{eq:merg_frac_c_corr}
\end{equation}
where $\mathrm{N_{BCGs,SC}}$ is the number of BCGs with spectroscopic companions and $C$ is the applied correction for spectroscopic incompleteness (derived in Appendix \ref{ap:corr_spec_comp}).

In our high redshift bin $0.4 \leq z \leq 0.5$, due to the low spectroscopic completeness of the SDSS at these redshifts, we used the SALT sample to determine the contamination correction (second term in Eq. \ref{eq:merg_frac}) i.e:
\begin{equation}
f_{\mathrm{pair}} \, = \, \frac{\mathrm{N_{BCGs,C}}}{\mathrm{N_{BCGs}}} \times \frac{\mathrm{N_{BCGs,BC(SALT)}}}{\mathrm{N_{BCGs, C(SALT)}}}
\label{eq:salt_bcgs_merger_frac}
\end{equation} 
We have assumed here that the fraction of potential merger pairs in the SALT sample is representative of those found in the photometric pair sample (selected from the GMBCG catalogue). The SALT pairs are not different from those in the photometric pair sample since the same $r_{\mathrm{sep}}$ and $\Delta m_{\mathrm{i}}$ criteria have been used to select pairs in both samples. The only difference between these two samples is the redshift requirement used for the SALT pair selection. Since the BCGs are bright, red galaxies, they are very likely to be observed by the SDSS spectroscopic surveys, so we do not expect a bias against spectroscopy for these galaxies. The number of pairs observed with SALT should therefore be representative of the number of pairs in the photometric sample.

The uncertainty on $f_{\mathrm{pair}}$ represents the 68.3 per cent $(1\sigma)$ binomial confidence limit and is calculated using the beta confidence interval as described by \cite{Cameron2011}.

\subsection{The mass growth of the BCGs}
\label{sec:mass_merger_growth}

We now continue on to the main aim of the paper, measuring the merger-inferred stellar mass growth of the BCGs as a function of redshift. In order to do this we first calculate average stellar masses of the BCGs, denoted $\langle M_{\star}\rangle$, in each redshift bin. The uncertainties on $\langle M_{\star}\rangle$ are determined using bootstrap resampling with 1000 realizations.

We estimate average merger timescale, denoted $\langle t_{\mathrm{merge}} \rangle$, for each close pair using the results of \citet{Kitzbichler2008}. Using the Millennium Simulation \citep{Springel2005}, they find that the average $t_{\mathrm{merge}}$ for galaxies with $r_{\mathrm{sep}} \leq 50$ kpc and $\Delta v \leq 300$ \kms\ can be given by the following:
\begin{equation}
\langle t_{\mathrm{merge}} \rangle = 2.2\,\mathrm{Gyr} \, \frac{r_{\mathrm{sep}}}{50 \, \mathrm{kpc}} \, \Big( \frac{M_{\star,\mathrm{com}}}{5.5 \times 10^{10} \, \mathrm{M_{\odot}}}\Big)^{-0.3}  \, \Big(1 + \frac{z}{8}\Big)
\label{eq:t_merge}
\end{equation}
where $M_{\star,\mathrm{com}}$ is the total stellar mass of the companion at its observed redshift (determined using {\tt kcorrect}), $r_{\mathrm{sep}}$ is the physical separation distance (in kpc) and $z$ is the photometric redshift of the cluster. The uncertainties on $\langle t_{\mathrm{merge}} \rangle$ are given by $1\sigma$ standard deviation and are propagated through to the final measured fractional mass growths. Various works have commented that the \cite{Kitzbichler2008} merger timescales are significantly longer than the estimate from dynamical friction or the orbital period \citep[e.g.][]{Conroy2007a, Bertone2009, Conselice2009a, Kauffmann2010, Lotz2011}. The uncertainties on $\langle t_{\mathrm{merge}} \rangle$ in all the redshift bins are large enough to encompass differences with other methods (see Table \ref{table:redM_mass_growth}).

The merger rate, denoted $R_{\mathrm{merge}}$, of the sample, i.e. the number of mergers per BCG per Gyr, is defined as follows:
\begin{equation}
R_{\mathrm{merge}} = \frac{f_{\mathrm{pair}}}{\langle t_{\mathrm{merge}} \rangle}
\label{eq:R_merge}
\end{equation}
The uncertainty on $R_{\mathrm{merge}}$ is determined using the standard propagation of errors. 

Semi-analytical models, for example \cite{Conroy2007, Puchwein2010, Laporte2013} have shown that $30-80$ per cent of the companion's stellar mass ends up in the ICL during a merger with the BCG. During a merger, we assume that 50 per cent of the companion's stellar mass is transfered to the BCG \citep[$f_{\mathrm{mass}} = 0.5$; also used by][]{Liu2009, Liu2015, Burke2013, Lidman2013}. To account for the possible range of $f_{\mathrm{mass}}$ values, we assign an uncertainty of 20 per cent to our assumed $f_{\mathrm{mass}}$ value of 50 per cent (i.e. $f_{\mathrm{mass}} = 0.5 \pm 0.2$), which is propagated through to the measured fractional mass growths. In order to calculate how much mass $(\Delta M)$ major mergers add to a BCG from redshift $z_{i}$ down to $z=0$, we use the following:
\begin{equation}
\Delta M \, (z = z_{i} - 0) \, = \, R_{\mathrm{merge}} \times T_{\mathrm{LB}} \times \langle M_{\star} \rangle_{\mathrm{com}} \, (z_{i}) \times f_{\mathrm{mass}} 
\label{eq:delta_M}
\end{equation}
where $T_{\mathrm{LB}}$ and $\langle M_{\star} \rangle_{\mathrm{com}}(z_{i})$ respectively gives the lookback time and the average stellar mass of the companions at $z_{i}$.

The fractional contribution $(F)$ made by major mergers to the stellar mass of a present day BCG, denoted $\langle M_{\star} \rangle_{\mathrm{BCG}} (z=0)$, is defined as follows:
\begin{equation}
F = \frac{\Delta M (z = z_{i} - 0)}{\langle M_{\star} \rangle_{\mathrm{BCG}} (z=0)}
\label{eq:mass_merge_growth}
\end{equation}
with
\begin{equation}
\langle M_{\star} \rangle_{\mathrm{BCG}} (z=0) = \langle M_{\star} \rangle_{\mathrm{BCG}} (z_{i}) + \Delta M (z = z_{i} - 0) + \Delta M_{\mathrm{other}}
\label{eq:mass_bcg_z0}
\end{equation}
where $\langle M_{\star} \rangle_{\mathrm{BCG}}(z_{i})$ is the average stellar mass of the BCGs at redshift $z_{i}$ and $\Delta M_{\mathrm{other}}$ accounts for other sources of mass accretion, i.e. star formation and minor mergers, from redshift $z_{i}$ down to $z=0$. We do not take this term into consideration because we are unable to estimate the contribution from minor mergers in this work, as we are incomplete for these systems\footnote{Between $0.08 \leq z \leq 0.20$, we are complete for mergers with mass ratios down to 1:6 (see Section \ref{sec:pair_selection} and panel c of Fig. \ref{fig:mag_lim_21_5}).}. Secondly, we expect the contribution from star formation to be negligible for these low redshift BCGs. It is rare to find BCGs in the local Universe that are actively forming stars \citep[less than 1 per cent; e.g.][]{Liu2012, Fraser-McKelvie2014}. These starbursts have been found to contribute only $\sim1-3$ per cent \citep[e.g.][]{Sarazin1983, Cardiel1995, Pipino2009, Liu2012, Loubser2016} to the stellar mass of the BCGs. The uncertainty on $F$ is determined using the standard propagation of errors.

\section{Results and Discussion}
\label{sec:results_disc}

\subsection{The redshift evolution of the pair fraction}
\label{sec:fpair_results}

\subsubsection{The redMaPPer pair fraction}
\begin{figure*}
\centering
\includegraphics[scale=0.5]{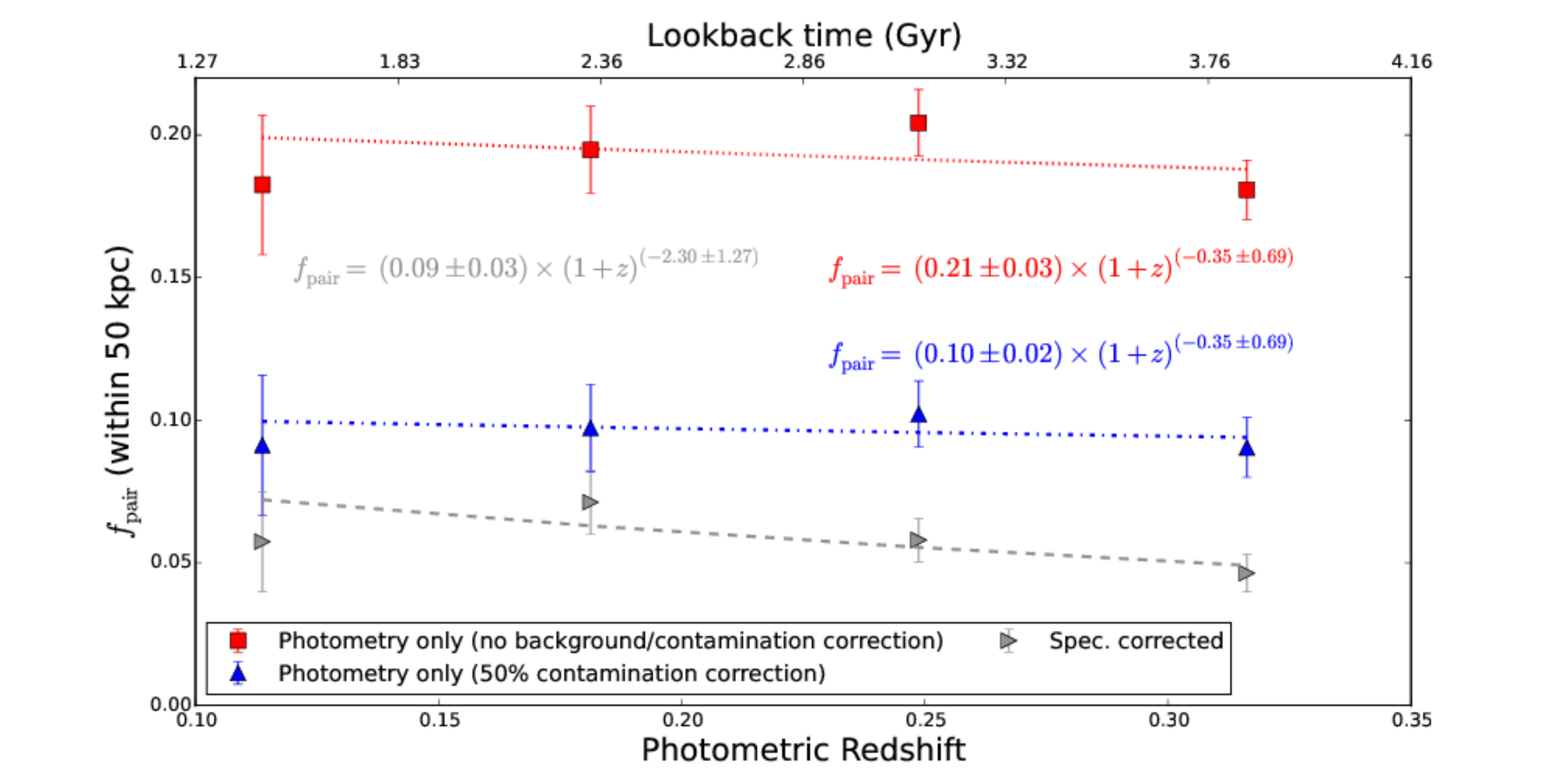}\\
\includegraphics[scale=0.5]{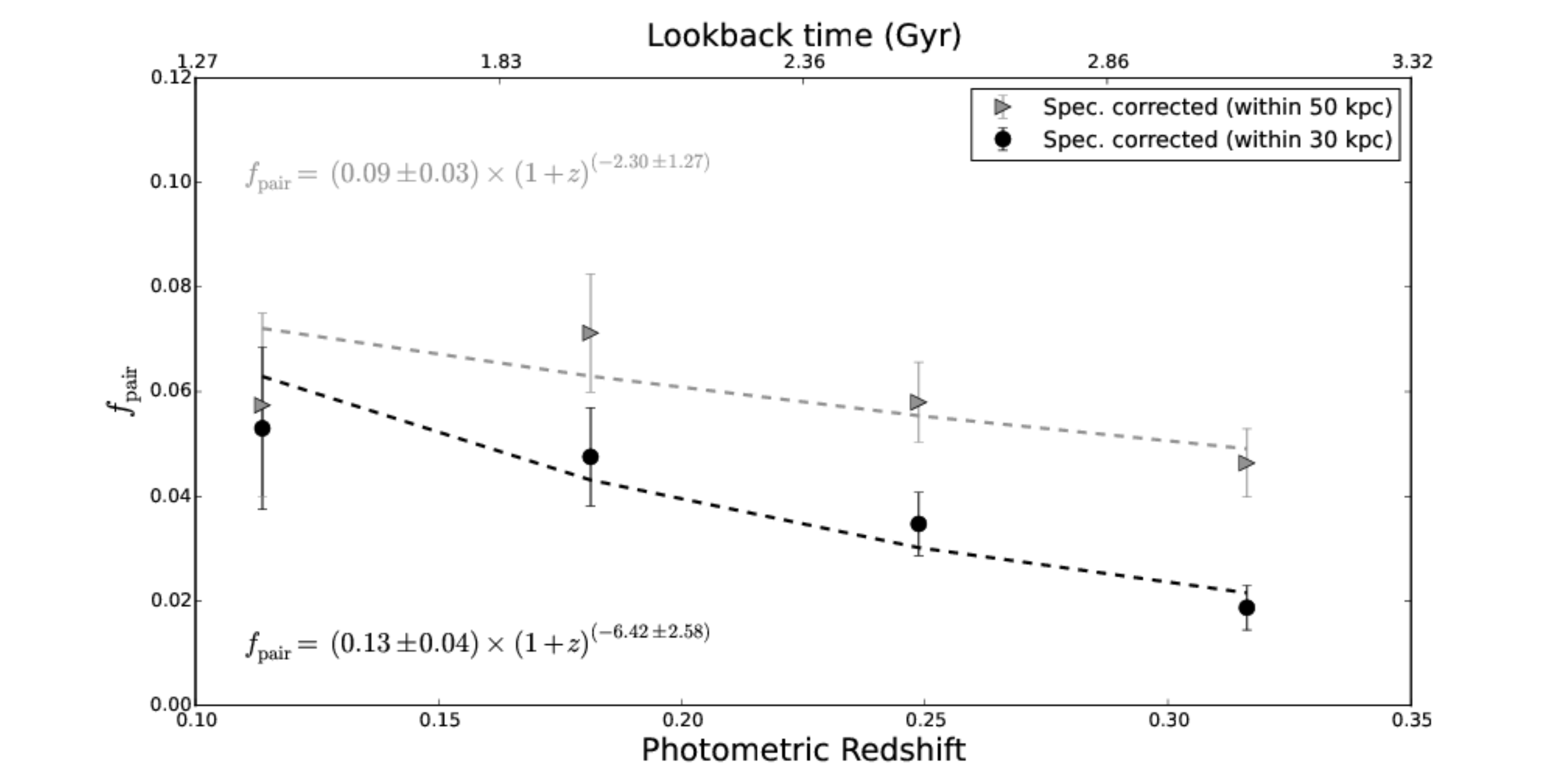}
\caption{The redshift evolution of the BCGs' pair fraction. In the top panel we show the photometric pair fraction derived within $7 \leq r_{\mathrm{sep}} \leq 50$ kpc. For comparison we also indicate $f_{\mathrm{pair}}$ if a 50 per cent contamination correction is applied and when the spectroscopic completeness is used to measure the pair fraction. In the bottom panel we show the comparison of the pair fractions that have been derived within 30 and 50 kpc (using the $C$-correction). Each line joins the pair fractions that have been obtained from the best-fitting power-law in each case. See text for details.}
\label{fig:fpair}
\end{figure*}

The redshift evolution of the pair fractions from $0.08 \leq z \leq 0.35$ for BCGs selected from the redMaPPer catalogue are shown in Fig. \ref{fig:fpair} and summarized in Table \ref{table:f_pair}. The pair fraction measured within 50 kpc using three different methods to correct for projection effects is shown in the top panel of Fig. \ref{fig:fpair}, while the bottom panel compares the spectroscopically-corrected results measured within 30 and 50 kpc.

A close pair sample constructed using only photometry with no background correction will inevitably suffer from projection effects. The resulting pair fraction can therefore be considered as an upper limit (red symbols and line in the top panel). To correct for this contamination, we have applied a 50 per cent contamination correction \citep[as used by][blue symbols and line]{Edwards2012}. We compare this pair fraction to that derived using the available spectroscopy, corrected for incompleteness as described above (grey symbols and line). We find that the spectroscopic $C$-correction reduces the photometric pair fraction to $\sim6$ per cent  while the simple assumption of 50 per cent contamination reduces the photometric pair fraction to $\sim10$ per cent.

Several studies in the literature, e.g. \cite{LeF`evre2000, Kartaltepe2007, 2009, Lopez-Sanjuan2012, Keenan2014} parameterised the evolution of the pair fraction with the following power-law function:
\begin{eqnarray}
f_{\mathrm{pair}} \, (z) = f_{\mathrm{pair}}(z=0) \times (1 + z)^{m}
\label{eq:fm_fit}
\end{eqnarray}

The power law fits to each of our three measurements are shown by the dashed and dotted lines in the figure. The negative power indices in each case suggest that the individual pair fractions increase slightly with decreasing redshift. The pair fraction derived using the spectroscopic $C$-correction has a steeper redshift evolution than photometric and the 50 per cent contamination-corrected pair fractions. 

The bottom panel of Fig. \ref{fig:fpair} compares the spectroscopic-corrected $f_{\mathrm{pair}}$ as measured within 30 kpc (black symbols and line) to that within 50 kpc. Here again, the negative power indices suggest that the pair fractions are increasing with decreasing redshift. The pair fraction within 50 kpc is a factor of $\sim2$ larger than that measured within 30 kpc. This is close to the expected difference in normalization for different radii found in previous works (e.g. from the two-point correlation function, \citealt[]{Bell2006a}). Although the observed evolution of $f_{\mathrm{pair}}$ is slightly steeper within the smaller radius, it is consistent within the uncertainty to that measured within the larger search radius of 50 kpc. 

Recall that the clusters in redMaPPer have been detected by looking for overdensities of red-sequence galaxies (Section \ref{sec:redM_over}). This may tempt the reader to think that the given cluster members are only red-sequence galaxies. If this is the case then we may miss potential merger candidates that are not located on the red-sequence, causing us to potentially underestimate the pair fraction of the BCGs. In Fig. \ref{fig:PR_rms_cuts_red_seq} we showed that galaxies with high probabilities $(P_{\rm{MEM}} > 0.9)$ form a tighter sequence, as expected, since they are more likely to be red-sequence galaxies. Those with smaller $P_{\rm{MEM}}$ values, on the other hand, are more scattered in colour. This suggests that some galaxies that are not on the red-sequence (i.e. galaxies in the blue cloud) are also included in the catalogue. We investigate this further using a method similar to that of \cite{Lu2009} to separate red-sequence galaxies from their blue counterparts. We refer the reader to \cite{Lu2009} for details. Briefly, we obtained the width of the red-sequence by fitting a single Gaussian against the galaxies with high probabilities of being red-sequence cluster members $(P_{\rm{MEM}} > 0.9)$. Galaxies within $2\sigma$ of the red-sequence are considered to be part of the red-sequence while those below this limit are classified as blue galaxies. We find that a small fraction (7 per cent) of the galaxies in redMaPPer are blue. Querying the SDSS DR8 database, we find an additional $\sim 200$ blue galaxies neighbouring our BCGs that are not included in the redMaPPer catalogue. Including these `missed' galaxies into our close pair sample and remeasuring the photometric pair fraction, results in an increase of less than one per cent. It is therefore clear that these `missing' blue galaxies have a negligible influence on the average BCG pair fraction.

Several studies in the literature use luminosity rather than stellar mass to identify potential major merger candidates \citep[e.g.][and others]{Liu2009, Liu2015, Edwards2012, Burke2013, Keenan2014, Burke2015}. We find that the selecting pairs based on their luminosity ratio rather than their mass ratio has no perceivable effect on our individually measured BCG pair fractions.

Recall that the pair fractions quoted here have been derived by only considering the BCGs that form part of the evolutionary cluster sequence (i.e. the halo masses of the clusters have been taken into account). We find these pair fractions to be consistent within the uncertainties to that measured in the case where we did not take the host clusters' $M_{h}$ growth into account. Additionally, we divide the cluster sample into a subsample of low and high mass clusters by using a fixed\footnote{This cut has been varied, however the results remain unchanged.} $M_{h}$ cut of $\sim 0.3 \times 10^{15} \, M_{\odot}$. The resulting pair fractions of these two subsamples are also consistent within the uncertainties. These results may suggest that the pair fraction of BCGs involved in major mergers does not depend on the halo mass of their host clusters.

\begin{table*}
\caption[]{Summary of the measured BCG pair fractions in the four redshift bins. Column 2 and 3 respectively state the $M_{h}$ range and number of clusters/BCGs that form part of the evolutionary sequence. Column 4 and 5 give the pair fraction obtained by only considering photometry (derived using Eq. \ref{eq:merg_frac}) and applying a 50 per cent contamination correction. The pair fractions given in Columns 6 and 7 are the spectroscopically corrected pair fractions (derived using Eq. \ref{eq:merg_frac_c_corr}). The uncertainties on $f_{\mathrm{pair}}$ have been derived using the method described by \cite{Cameron2011}. We indicate the physical separation distance within which these $f_{\mathrm{pair}}$ have been measured.}
\label{table:f_pair}
\begin{tabular}{ccccccc} 
		\hline
		
        \multicolumn{1}{c}{} &
        \multicolumn{1}{c}{} &
        \multicolumn{1}{c}{} &        
		\multicolumn{4}{c}{$f_{\mathrm{pair}}$} \\
        
		\multicolumn{1}{c}{$z$} &
        \multicolumn{1}{c}{$M_{h}$ range} &
        \multicolumn{1}{c}{$N_{\mathrm{BCG}}$} &
        \multicolumn{1}{c}{Phot. only} &
		\multicolumn{1}{c}{50\% contamination corr.} &
        \multicolumn{2}{c}{Spec. corr.} \\

        \multicolumn{1}{c}{} & 
        \multicolumn{1}{c}{($\times 10^{15} \, \mathrm{M_{\odot}}$)} &
        \multicolumn{1}{c}{} & 
		\multicolumn{1}{c}{(50 kpc)} &
		\multicolumn{1}{c}{(50 kpc)} &
        \multicolumn{1}{c}{(50 kpc)} &
        \multicolumn{1}{c}{(30 kpc)} \\
        
		\hline
		
		$0.08 - 0.15$ & $> 0.29$ & 329 & $0.183 \pm 0.025$ & $0.091 \pm 0.012$ & $0.057 \pm 0.018$ & $0.053 \pm 0.015$ \\
		$0.15 - 0.21$ & $> 0.26$ & 789 & $0.195 \pm 0.015$ & $0.097 \pm 0.008$ & $0.071 \pm 0.011$ & $0.048 \pm 0.009$ \\
		$0.21 - 0.28$ & $> 0.24$ & 1572 & $0.204 \pm 0.012$ & $0.102 \pm 0.006$ & $0.058 \pm 0.008$ & $0.035 \pm 0.006$ \\
		$0.28 - 0.35$ & $> 0.22$ & 2742 & $0.181 \pm 0.010$ & $0.090 \pm 0.005$ & $0.046 \pm 0.007$ & $0.019 \pm 0.004$ \\
		\hline
\end{tabular}
\end{table*}

\subsubsection{The SALT pair fraction}

In order for a close galaxy pair to be considered a potential merger candidate, we require $r_{\mathrm{sep}} \leq 50$ kpc and $\Delta v \leq 300$ \kms. We find that seven of the 12 close pairs in the SALT sample satisfy this criteria (see Table \ref{table:SALT_sample}). Using Eq. \ref{eq:salt_bcgs_merger_frac}, we find the measured bound BCG pair fraction to be $0.21^{+0.04}_{-0.05}$ between $0.4 \leq z \leq 0.5$. If instead we consider a separation radius of 30 kpc, we find that three of the pairs in the SALT sample can be considered potential merger candidates. This results in a bound BCG pair fraction of $0.09^{+0.06}_{-0.03}$ (as derived using Eq. \ref{eq:salt_bcgs_merger_frac}).

\subsubsection{Literature comparison}

\begin{figure*}
\centering
\includegraphics[scale=0.5]{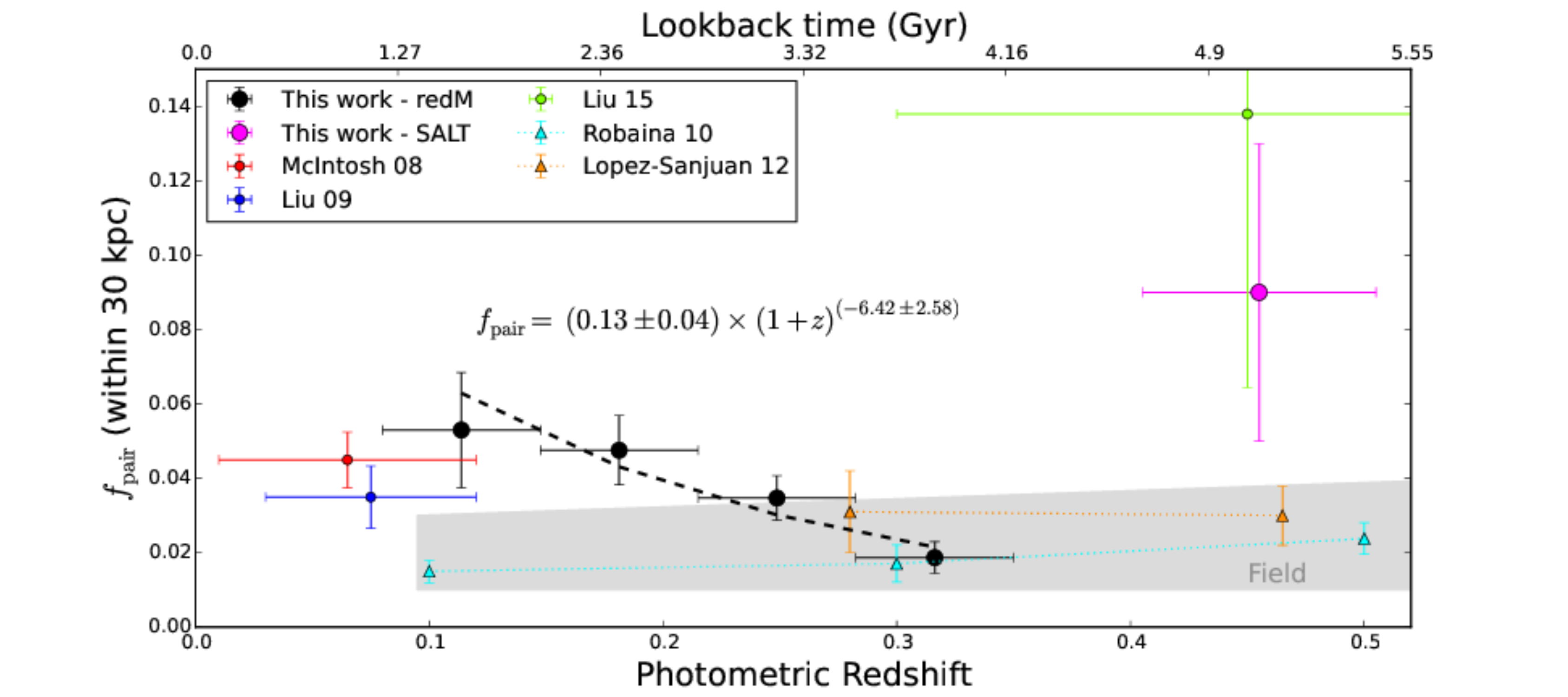}
\caption[]{Literature comparison of the major merger pair fraction within $r_{\mathrm{sep}} = 30$ kpc. The dashed line joins the pair fractions that have been obtained from the best-fitting power-law. The points indicate values obtained from studies done in clusters while the shaded region represents the values obtained from studies done in the field. As examples we only indicate the pair fractions of two (represented by the triangles) of those field studies. The pair fractions from the literature are plotted at the average redshifts of the samples while the horizontal error bars indicate the width of the redshift bins. See text for details.}
\label{fig:fpair_lit_comp_30kpc_evolB_SALT}
\end{figure*}

In Fig. \ref{fig:fpair_lit_comp_30kpc_evolB_SALT} we show the comparison of our measured pair fraction against values obtained from the literature. We compare against the results of studies conducted in clusters and the field. The majority of the studies discussed hereafter have derived pair fractions within $r_{\mathrm{sep}} \leq 30$ kpc. We have therefore chosen to only show the spectroscopically corrected pair fraction we obtained by considering close pairs within the same physical separation distance (black points).

\subsubsection*{Cluster studies:}

Both \cite{McIntosh2008} and \cite{Liu2009} have studied the BCG pair fraction (over $0.01 \leq z \leq 0.12$) by searching for close pairs within 30 kpc. They further restricted their samples to only consider close pairs where both the BCG and companion show signs of morphological distortions (i.e. diffuse tails and asymmetries in the inner isophotes). These distortions indicate that the galaxies are in the process of merging. By only considering major mergers (luminosity ratios $\leq$ 1:4), \cite{McIntosh2008} found that 38 of their 845 close pairs were morphologically distorted, giving $f_{\mathrm{pair}} = 0.04 \pm 0.01$, assuming Poisson errors (red point). Similarly, \cite{Liu2009} found $f_{\mathrm{pair}} = 0.03 \pm 0.01$ (18/515) by assuming Poisson errors (blue point).

Only our first redshift bin overlaps with the redshift range used in the above-mentioned studies $(z \leq 0.12)$. We measure a pair fraction of $0.05 \pm 0.01$ at $z\sim 0.11$. The main difference between our study and that of \cite{McIntosh2008,Liu2009} is the technique used to select mergers: morphological distortions vs. close pairs. Mergers identified through morphological distortions are in the final stages of merging. Galaxies in these pairs are typically expected to merge within $\sim0.2$ Gyr \citep[e.g.][]{Patton2002, Hernandez-Toledo2005, Lotz2011}. The close pair technique on the other hand selects mergers that are in various stages of the merging process (early to final stages). For major merger pairs with a separation distance of $\sim 50$ kpc, \cite{Lotz2011} estimate a merger timescale of roughly 0.6 Gyr. For pairs with $r_{\mathrm{sep}} \sim 30$ kpc, $t_{\mathrm{merge}}$ decreases to 0.33 Gyr \citep{Lotz2011}. The merger timescales of the pairs in our sample are therefore longer (on average) than those in \cite{McIntosh2008,Liu2009}. It is therefore not unreasonable that we find a higher pair fraction. 

We are not able to measure the three dimensional velocities of galaxies so we can only use their line-of-sight velocities from spectroscopy. By imposing a velocity difference (as done in our analysis), one is able to identify pairs of galaxies that are potentially gravitationally bound to one another. False-positive pairs may still be included in the sample, despite the velocity cut, making our pair fraction potentially higher than \cite{McIntosh2008,Liu2009}. The velocity cut on the other hand may also exclude real bound pairs from the sample, if they have larger velocity differences than used in this work. However, when  we use $\Delta v \leq 500$ \kms\ to identify potential merger candidates \citep[also used by][]{Lin2004a, Lin2008, Lopez-Sanjuan2012, Robotham2014}, we find that it only increases $f_{\mathrm{pair}}$ by $\sim 0.03$ per cent. Thus we find that the exclusion of real bound pairs, which do not satisfy our velocity cut, has a negligible effect on the measured pair fraction. 

In a later study, \cite{Liu2015} extended their analysis to investigate the role of major mergers in the mass growth of BCGs between $0.3 \leq z \leq 0.6$ (with a median value of $z\sim 0.43$). Using the same methodology as \cite{Liu2009}, they find that 4 of their 29 BCGs have morphological distortions, giving $f_{\mathrm{pair}} = 0.14 \pm 0.07$, assuming Poisson errors (green point). 

Using data from SALT we measure a bound pair fraction of $0.09 ^{+0.06}_{-0.03}$ at $z=0.45$ within 30 kpc (magenta point). This is lower than that found by \cite{Liu2015}, however the pair fractions are consistent within the uncertainties.

\subsubsection*{Field studies:}

There are numerous studies over the years that have investigated the close pair fraction of galaxies in the field \citep[e.g.][]{Carlberg2000, Patton2000, Lin2004a, Lin2008, Kartaltepe2007,  Hsieh2008, Bundy2009, 2009, Jogee2009, Robaina2010, Lopez-Sanjuan2012, Keenan2014, Robotham2014}. The shaded region in Fig. \ref{fig:fpair_lit_comp_30kpc_evolB_SALT} represents the range of pair fractions that are measured in the field. We only indicate the pair fractions of two field studies as examples. They were chosen because they defined major mergers in the same way as we do in this paper.

Most studies of the pair fraction in the field have been for galaxies of approximately $L^{\star}$ rather than the very massive BCGs we consider here. Nevertheless, it is useful to compare the range of values found in the field to that measured in clusters. This is done to determine how the redshift evolution of the major merger pair fractions in these two environments compare. 

\citet{Jogee2009, Keenan2014, Robotham2014} investigated the close pair fraction of field galaxies with $M_{\star} \geq (2 - 5) \times 10^{10} \, \mathrm{M_{\odot}}$ over $0.1 \leq z \leq 1.2$. The close pairs were selected to be within $r_{\mathrm{sep}} \leq 30$ kpc and $\Delta v \leq 500$ \kms, although \cite{Keenan2014} also imposed a lower $r_{\mathrm{sep}}$ limit of 7 kpc on their close pair sample selection. These studies all considered major mergers, although the major merger definition differed slightly from study to study. \cite{Jogee2009} used stellar mass ratios of (1:4), while \cite{Robotham2014} used a mass ratio of (1:3). \cite{Keenan2014} in turn used luminosity ratios of (1:4) to define major mergers. The results from these studies suggested that the pair fraction experienced a very mild evolution since $z\sim1$. 

\cite{Robaina2010} on the other hand measured the pair fraction (within 30 kpc) of field galaxies with $M_{\star} > 5 \times 10^{10} \, \mathrm{M_{\odot}}$ by using the amplitude of the two-point correlation function. Although they do not impose a specific mass ratio criteria, they find that most of their merger sample ($\sim 90$ per cent) consists of major mergers (i.e. mass ratios between 1:1$-$1:4). They measure a very mild evolution in the pair fraction since $z\sim1$.

The studies of \cite{Lin2008, Lopez-Sanjuan2012} used $10 \lesssim r_{\mathrm{sep}} \lesssim 50$ kpc and $\Delta v \leq 500$ \kms\ to identify close galaxy pairs in the field between $0.2 \leq z \leq 0.9$. The galaxies in these studies had $M_{\star} \geq 10^{11} \, \mathrm{M_{\odot}}$. Both these studies only considered major mergers, however \cite{Lopez-Sanjuan2012} used stellar mass ratios of (1:4) to identify major mergers, while \cite{Lin2008} used luminosity ratios of (1:4) instead. These studies also separated the red and blue galaxies in their samples to investigate the pair fraction for each galaxy population. \cite{Lopez-Sanjuan2012} found that the major merger rate for blue galaxies decreased rapidly with decreasing redshift, while this evolution was slower for red galaxies. \cite{Lin2008} observed the same trend for their blue galaxies, however they found that the red galaxy pair fraction increased with decreasing redshift. The same results were observed by other studies that separated red (or early-type) and blue (or late-type) galaxies \citep[e.g.][]{Bundy2009, 2009}. More interesting though is that they found that the major merger fraction of early-type galaxies was higher than that of late-type galaxies. \cite{Robotham2014} found a similar result, i.e. the major merger pair fraction of high mass galaxies was higher than for lower mass galaxies. \cite{Liu2009} in turn found that the pair fraction (specifically for major mergers) increased with halo mass.

Although a direct comparison between our observed $f_{\mathrm{pair}}$ and that measured in the field is beyond the scope of this work, we find that all of the above-mentioned field studies indicate that little to no evolution took place in $f_{\mathrm{pair}}$ since $z \sim 1$. Our results on the other hand suggest that the major merger pair fraction of BCGs (in clusters, by definition) increases with decreasing redshift, a result that was also found by \cite{Lin2008} for the red galaxies in their field sample.

\subsection{Stellar mass growth due to mergers}

In Table \ref{table:redM_mass_growth} we present a summary of the fractional contribution made by major mergers to the stellar mass of a present day BCG since a given redshift (assuming an evolving merger rate). Recall that these results have been derived under the assumption that half of the stellar mass of the companion is accreted onto the BCG $(f_{\mathrm{mass}} = 0.5 \pm 0.2)$. We find that major mergers contributed $24 \pm 14$ per cent (on average) to the stellar mass of a present day BCG since $z=0.32$. When we also take the SALT sample into consideration, we find the fractional contribution made by major mergers to the stellar mass of a present day BCG to be $29 \pm 19$ per cent, on average, since $z=0.45$. 

We note that since the merger rate is very constant with redshift (see Table \ref{table:redM_mass_growth}), the mass growth results are very similar if we integrate Eq. \ref{eq:delta_M} using a fit to the measured merger rates or assume the rate is constant in each redshift bin, given by the rate at the centre of that bin.

\begin{table}
\caption{Summary of the fractional contribution (Column 4) made by major mergers to the stellar mass of a present day BCG since a given redshift. The average merger timescales and merger rates at the given redshifts are also indicated in Columns 2 and 3. The uncertainties have been determined using the standard propagation of errors.}
\label{table:redM_mass_growth}
\begin{tabular}{cccl} 
		\hline
		
		\multicolumn{1}{c}{$z$} &
		\multicolumn{1}{c}{$\langle t_{\mathrm{merge}} \rangle$} &
        \multicolumn{1}{c}{$R_{\mathrm{merge}}$} &
        \multicolumn{1}{c}{$F$} \\

		\multicolumn{1}{c}{} &
        \multicolumn{1}{c}{(Gyr)} &
		\multicolumn{1}{c}{(Gyr$^{-1}$)} & 
        \multicolumn{1}{c}{(per cent)} \\
		
		\hline
		
		\multicolumn{4}{c}{\textit{redMaPPer:}} \\
		0.11 & $0.79 \pm 0.53$ & $0.07 \pm 0.06$ & $13 \pm 8$  \\
		0.18 & $0.69 \pm 0.52$ & $0.10 \pm 0.08$ & $20 \pm 12$ \\
		0.25 & $0.63 \pm 0.50$ & $0.09 \pm 0.08$ & $24 \pm 15$ \\
		0.32 & $0.77 \pm 0.55$ & $0.06 \pm 0.05$ & $24 \pm 14$ \\
		
		\multicolumn{4}{c}{ } \\
		\multicolumn{4}{c}{\textit{SALT:}} \\
		0.45 & $0.89 \pm 0.42$ & $0.24 \pm 0.12$ & $29 \pm 17$\\
		\hline
\end{tabular}
\end{table}

\subsubsection{Literature comparison}

\begin{figure*}
\centering
\includegraphics[scale=0.5]{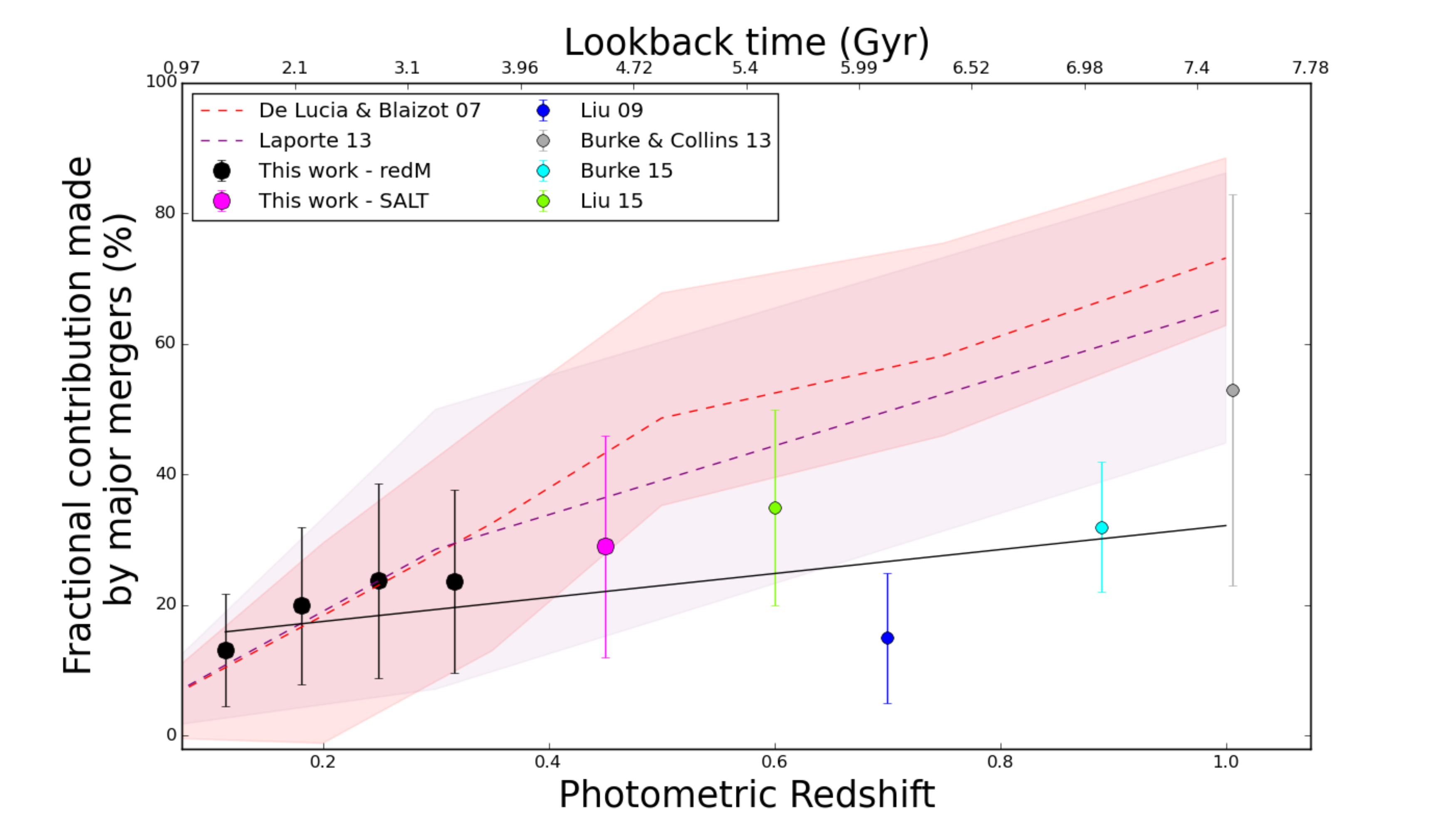}
\caption[]{Comparison of our measured fractional contribution made by major mergers to the stellar mass of a present day BCG since a given redshift. The fractional contributions have been measured from the redshift they are plotted at, down to $z=0$ (assuming $f_{\mathrm{mass}} = 0.5$). The studies shown as points are observational studies while model predictions are indicated with the shaded regions. The solid and dashed lines respectively indicate the best-linear fit (obtained using least-square fitting) to the observational studies and simulations. See text for details.}
\label{fig:fracmassg_lit_cumm_salt}
\end{figure*}

In Fig. \ref{fig:fracmassg_lit_cumm_salt} we show the comparison of our measured BCG mass growth (due to major mergers within $7 \leq r_{\mathrm{sep}} \leq 50$ kpc) to that derived from studies in the literature. We only show the mass growth results (due to major mergers) that have been derived under the assumption that half of the companion's mass is accreted onto the BCG. The solid and dashed lines respectively indicate the best-linear fit, obtained using least-square fitting, to the observational studies (represented by the filled circles) and simulations (represented by the shaded regions). 

Using morphological distortions as an indication of mergers, \cite{Liu2009} found that major mergers would contribute $15 \pm 10$ per cent on average to the stellar mass of a present day BCG from $z=0.7$ (blue point) at a mean rate of $2.5 \pm 1.7$ per cent per Gyr (assuming $f_{\mathrm{mass}} = 0.5$). \cite{Liu2015} found that major mergers, on average, contributed $35 \pm 15$ per cent to the stellar mass of a present BCG since $z=0.6$ by assuming $f_{\mathrm{mass}} = 0.5$ (green point).

In a separate study, \cite{Burke2015} used close pairs (identified within 50 kpc) to study the stellar mass build-up of 23 BCGs over $0.18 < z < 0.90$. Their sample included mergers down to stellar mass ratios of 1:20 and therefore considered both major and minor mergers. They found that these mergers increased the BCGs' stellar masses by a factor of 1.2 since $z = 0.9$. They estimate that $32 \pm 10$ per cent of the total merging mass is locked up in major mergers (with mass ratios of 1:2$-$1:5, cyan point). Extending this analysis out to $0.8 < z < 1.4$, \cite{Burke2013} found that major mergers (mass ratios of 1:1$-$1:2) contributed $53 \pm 30$ per cent to the stellar mass of present day BCGs since $z\sim1$ (grey point), while the remainder was contributed by minor mergers (mass ratios of $>$1:3$-$1:20).

At $z\sim1$ the \citet{DeLucia2007, Laporte2013} semi-analytical models (represented by the red and purple dashed lines) predict larger mass growth results for the BCGs than what is observed, with the SAM of \cite{DeLucia2007} predicting the largest growth. At lower redshifts, however, both these models predict mass growth results that are in agreement with the observations. There may be two sources responsible for the discrepancy at higher redshifts. The first is that the SAMs consider both major and minor mergers. The second source may lie in the manner in which the SAMs take the efficiency of mergers into account, specifically the fraction of the accreted mass that ends up in the ICL. The SAM of \cite{Laporte2013} predicts a mass growth that is in better agreement with what is observed, perhaps because they include an ICL prescription, with 30 per cent of the companion mass transferred to the ICL in each merger. \cite{DeLucia2007} does not account for loss to the ICL. This illustrates the need to take the stellar mass growth of the ICL into account when the BCG stellar mass growth is investigated.

\subsection{Contribution of mergers towards the stellar mass build-up of the ICL}

Galaxies merging with BCGs have been proposed as a mechanism through which the stellar mass of the ICL grows. Tidal stripping of satellite galaxies are also thought to contribute towards the stellar mass growth of the ICL. This idea is taken into consideration by the more recent simulations that are used to study ICL formation \citep[e.g.][]{Contini2014, DeMaio2015}. The simulations can be used study the mass assembly of BCGs and taking the tidal stripping of galaxies into account during the ICL formation may serve to alleviate the discrepancies found between the predicted and observed stellar mass growth of the BCGs. Observations and simulations of the ICL suggest that the stellar mass of the ICL has increased by a factor of $4-5$ since $z=1$ \citep[e.g.][]{Krick2007, Murante2007, Rudick2011, Burke2012, Burke2015}. The N-body simulations of \cite{Murante2007, Conroy2007, Puchwein2010} predict that $30-80$ per cent of the BCGs' merging mass should be distributed into the ICL in order to reproduce the distribution observed in nearby clusters. Combined with the number of mergers that BCGs are expected to experience \citep{Edwards2012, Burke2013, Lidman2013}, one is faced with a scenario where the ICL is being built up through galaxies that are interacting with BCGs. If we are to believe the model predictions that much of the mass from companions that merge with BCGs is distributed into the ICL, then this could explain why the stellar masses of the BCGs remain relatively unchanged (or only slightly increase) with decreasing redshift. We investigate whether major mergers provide sufficient stellar material to explain the stellar mass growth of the ICL from $z=0.3$ and $z=0$. We denote the amount of stellar mass added to the ICL due to mergers as $(\Delta M_{\star})_{\mathrm{mergers}}$ and the stellar mass growth of the ICL between two redshift bins by $(\Delta M_{\star})_{\mathrm{ICL}}$. If $(\Delta M_{\star})_{\mathrm{ICL}} \gg (\Delta M_{\star})_{\mathrm{mergers}}$, it implies that mergers do not provide enough stellar mass to account for the growth of the ICL. We investigate this idea by combining values from the literature with our own measurements.

We were not able to obtain a stellar mass estimate of the ICL at either $z=0.3$ or $z=0$ from the literature, however \cite{DeMaio2015} investigated the ICL stellar mass build-up in four clusters at $0.44 \leq z \leq 0.57$ (with a median redshift of $z=0.5$) that span a halo mass range of $0.6-2.6 \times 10^{15} \, \mathrm{M_{\odot}}$. They measured the ICL luminosities of these clusters within a radius of 110 kpc from their centres (see their table 6). Because their $M_{h}$ range overlaps with ours, we take the average of their measured ICL luminosities, denoted $\langle L^{\star} \rangle_{\mathrm{ICL}}(z=0.5)$, as a proxy of the ICL's average stellar mass at $z=0.5$, denoted $\langle M_{\star} \rangle_{\mathrm{ICL}}(z=0.5)$. We then use the  predicted stellar mass growth factor $(f)$ of the ICL from \citet[][see their figure 6]{Contini2014}\footnote{They investigated the ICL growth in clusters by updating the \cite{DeLucia2007} semi-analytical model to include various implementations for the formation of the ICL. Specifically, the ICL is assumed to form from the stellar stripping of satellite galaxies and mergers with the BCGs (assuming that 20 per cent of the companion's stellar mass ends up in the ICL during a merger event with the BCG).} to determine $\langle M_{\star} \rangle_{\mathrm{ICL}}(z=0)$ and $\langle M_{\star} \rangle_{\mathrm{ICL}}(z=0.3)$ as described below. 

\begin{eqnarray}
(\Delta M_{\star})_{\mathrm{ICL}} = \langle M_{\star} \rangle_{\mathrm{ICL}}(z=0) - \langle M_{\star} \rangle_{\mathrm{ICL}}(z=0.3)
\label{eq:ICL_mass_growth_1}
\end{eqnarray}
with 
\begin{align}
\begin{split}
\langle M_{\star} \rangle_{\mathrm{ICL}}(z=0) & = \frac{\langle M_{\star} \rangle_{\mathrm{ICL}}(z=0.5)}{f(z=0.5-0)}
\\
& = \frac{\langle L^{\star} \rangle_{\mathrm{ICL}}(z=0.5) \times (M_{\star})_{L^{\star}}}{f(z=0.5-0)} 
\label{eq:ICL_mass_0}
\end{split}
\end{align}
where $\langle L^{\star} \rangle_{\mathrm{ICL}}(z=0.5)$ denotes the average ICL luminosity of the clusters at $z=0.5$ and has a value of $7.4 \, L^{\star}$ \citep[obtained from][]{DeMaio2015}. The stellar mass of a $L^{\star}$ galaxy, denoted $(M_{\star})_{L^{\star}}, is \sim 5 \times 10^{10} \, \mathrm{M_{\odot}}$.

\begin{eqnarray}
\langle M_{\star} \rangle_{\mathrm{ICL}}(z=0.3) = \langle M_{\star} \rangle_{\mathrm{ICL}}(z=0) \, \times \, f(z=0.3-0)
\label{eq:ICL_mass_growth_3}
\end{eqnarray}
where $f(z=0.5-0) = 0.5$ and $f(z=0.3-0) = 0.75$. These are the predicted stellar mass growth factors of the ICL between $z=0.5-0$ and $z=0.3-0$ from \cite{Contini2014}. Substituting these values into the relevant equations above, we find $(\Delta M_{\star})_{\mathrm{ICL}} \sim 2 \times 10^{11} \, \mathrm{M_{\odot}}$.

We measure the amount of stellar mass contained in the close companions of BCGs to be $\sim 4 \times 10^{11} \, \mathrm{M_{\odot}}$, of which half is transferred to the ICL. Compared to the ICL stellar mass estimate given above, we find there is sufficient stellar material in the galaxies that are likely to merge with the BCGs as major mergers to account for the stellar mass growth of the ICL between $0.0 \leq z \leq 0.3$. This may also imply that our assumption of 50 per cent mass transfer to the BCG during mergers is enough to grow the ICL, arguing against a higher mass transfer. 

\section{Conclusions}
\label{sec:concl}

In this paper, we examine the role that mergers play in the stellar mass build-up of BCGs between $0.08 \leq z \leq 0.50$. For this purpose we identify close galaxy pairs within a 50 kpc physical search radius and assume that these close pairs are diagnostics of mergers. The close pair fraction is then used to determine how much stellar mass growth the BCGs have experience over the redshift range of interest due to major mergers. 

\begin{itemize}
\item We observe a weak trend that the spectroscopically corrected BCG pair fraction increases with decreasing redshift, suggesting that major mergers may become more important towards the present day. The evolution of the pair fractions (Eq. \ref{eq:fm_fit}) within 30 and 50 kpc are respectively given by $m = -6.42 \pm 2.58$ and $m = -2.30 \pm 1.27$. 
\item Since $z=0.32$ we find the fractional contribution made by major mergers to the stellar mass of a present day BCG to be $24 \pm 14 $ per cent, on average. 
\item Using data from SALT we extend our study to $z=0.45$. From this redshift, we find the fractional contribution made by major mergers to the stellar mass of a present day BCG to be $29 \pm 17$ per cent, on average.
\item We also investigate whether mergers provide sufficient stellar material to explain the stellar mass growth of the ICL from $z=0.3$ to $z=0$. Using the predicted stellar mass growth factor from \cite{Contini2014} we find that the ICL has increased its stellar mass by $(\Delta M_{\star})_{\mathrm{ICL}} \sim 2 \times 10^{11} \, \mathrm{M_{\odot}}$ between $z=0.3$ and 0. From our analysis of the amount of mass contained within the close companions of BCGs, we estimate that major mergers distributed about $(\Delta M_{\star})_{\mathrm{mergers}} \sim 2 \times 10^{11} \, \mathrm{M_{\odot}}$ of stellar mass into the ICL between $z=0.3$ to $z=0$. Our findings imply that galaxies that are likely to merge with the BCG as major mergers provide enough stellar material to account for the stellar mass growth of the ICL (at least from $z \leq 0.3$).
\end{itemize}


\section*{Acknowledgements}

We thank the anonymous referee for the many helpful suggestions that improved this paper. The authors gratefully acknowledge the financial support of the National Research Foundation (NRF) towards this project. DNG and RES are funded under the Professional Development Programme by the NRF.

DNG would like to thank Risa Wechsler, Danilo Marchesini and Claire Burke for helpful discussions. 

Funding for SDSS-III has been provided by the Alfred P. Sloan Foundation, the Participating Institutions, the National Science Foundation, and the U.S. Department of Energy Office of Science. The SDSS-III web site is \href{http://www.sdss3.org/}{http://www.sdss3.org/}.

SDSS-III is managed by the Astrophysical Research Consortium for the Participating Institutions of the SDSS-III Collaboration including the University of Arizona, the Brazilian Participation Group, Brookhaven National Laboratory, University of Cambridge, Carnegie Mellon University, University of Florida, the French Participation Group, the German Participation Group, Harvard University, the Instituto de Astrofisica de Canarias, the Michigan State/Notre Dame/JINA Participation Group, Johns Hopkins University, Lawrence Berkeley National Laboratory, Max Planck Institute for Astrophysics, Max Planck Institute for Extraterrestrial Physics, New Mexico State University, New York University, Ohio State University, Pennsylvania State University, University of Portsmouth, Princeton University, the Spanish Participation Group, University of Tokyo, University of Utah, Vanderbilt University, University of Virginia, University of Washington, and Yale University.

This work is based in part on observations taken with the Southern African Large Telescope (SALT), proposals 2013-2-RSA-008 and 2014-1-RSA\_OTH-009. We thank the South African Astronomical Observatory for allocating us the time. 

IRAF is distributed by the National Optical Astronomy Observatories, which are operated by the Association of Universities for Research in Astronomy, Inc., under cooperative agreement with the National Science Foundation.



\appendix

\section{Investigating the spectroscopic completeness of the close pair sample}
\label{ap:corr_spec_comp}

Not every galaxy in the close pair sample has a spectroscopic redshift. This may simply be because the galaxies did not meet the selection criteria of the SDSS's spectroscopic surveys or fiber collision problems caused them not to be targeted for spectroscopy. The minimum separation between the SDSS spectroscopic fibers is $55\arcsec$ and if two objects are closer than this, then only one will be observed \citep{Zehavi2002, Blanton2003}. Repeat observations however mean that many galaxies closer than $55\arcsec$ are observed in practice.   

Our close pair sample is therefore spectroscopically incomplete, which in turn will cause close pairs to be missed because one or more galaxies in these pairs have no spectroscopy. We correct for spectroscopic incompleteness by determining how the completeness of our close pair sample changes as a function of the selection criteria used in the SDSS spectroscopic galaxy samples (summarised in Table \ref{table:spec_sample_selec}). The angular separation ($r_{\mathrm{ang}}$) between the galaxies in the close pairs also needs to be considered since fiber collisions also influence the spectroscopic completeness of the close pair sample. Through this approach we assign each galaxy in the close pair sample a spectroscopic weight, which is used to determine the pair fraction.

\begin{table}
\caption{Summary of the selection criteria used in the MGS, LRG and BOSS sample to target galaxies for spectroscopy. Column 2 and 3 give the magnitude limit and galaxy colour selection of each survey respectively. Due to the transition of the 4000 \AA\ break from the \textit{g} to \textit{r}-band at $z\sim0.4$, both the LRG and BOSS surveys have to use two different selection criteria to select galaxies for spectroscopic follow-up below and above this redshift.}
\label{table:spec_sample_selec}
\begin{tabular}{cccc} 
		\hline
		
		\multicolumn{2}{c}{Sample} &
		\multicolumn{1}{c}{Magnitude} &
		\multicolumn{1}{c}{Colour}\\
		\multicolumn{2}{c}{} &
		\multicolumn{1}{c}{(mag)} &
		\multicolumn{1}{c}{} \\
		
		\hline
		
		MGS & & $r_{\mathrm{petro}} < 17.77$ & \\ 

		LRG & Cut I ($z<0.4$) & $r_{\mathrm{petro}} \leq 19.2$ & $(g - r)$ \\ 
		& Cut II ($z>0.4$) & $r_{\mathrm{petro}} \leq 19.5$ & $(r - i)$ \\ 

		BOSS & LOWZ ($z<0.4$)& $r_{\mathrm{cmod}} \leq 19.6$ & $(g - r)$ \\ 
		& CMASS ($z>0.4$) & $i_{\mathrm{cmod}} \leq 19.9$ & $(r - i)$ \\ 
		\hline
\end{tabular}
\end{table}

The correction for spectroscopic incompleteness (hereafter $C$-correction) we apply to the close pair sample is determined as a function of four variables: the $r$-band Petrosian magnitude ($r_{\mathrm{petro}}$), $(g - r)$ and $(r - i)$ colours of the galaxies along with the pairs' angular separation distances, $r_{\mathrm{ang}}$. The $C$-correction is given as:
\begin{eqnarray}
C\, = \frac{N_{\mathrm{spec. \,pairs}}(r_{\mathrm{petro}}, \mathit{g - r}, \mathit{r - i}, r_{\mathrm{ang}}) }{N_{\mathrm{phot. \,pairs}}(r_{\mathrm{petro}}, \mathit{g - r}, \mathit{r - i}, r_{\mathrm{ang}})} 
\label{eq:spec_comp_fac}
\end{eqnarray}
where $N_{\mathrm{spec. \,pairs}}$ and $N_{\mathrm{phot. \,pairs}}$ respectively indicates the number of pairs in the spectroscopic and photometric samples. 

\begin{figure*}
\centering
\includegraphics[width=15cm, height=14cm]{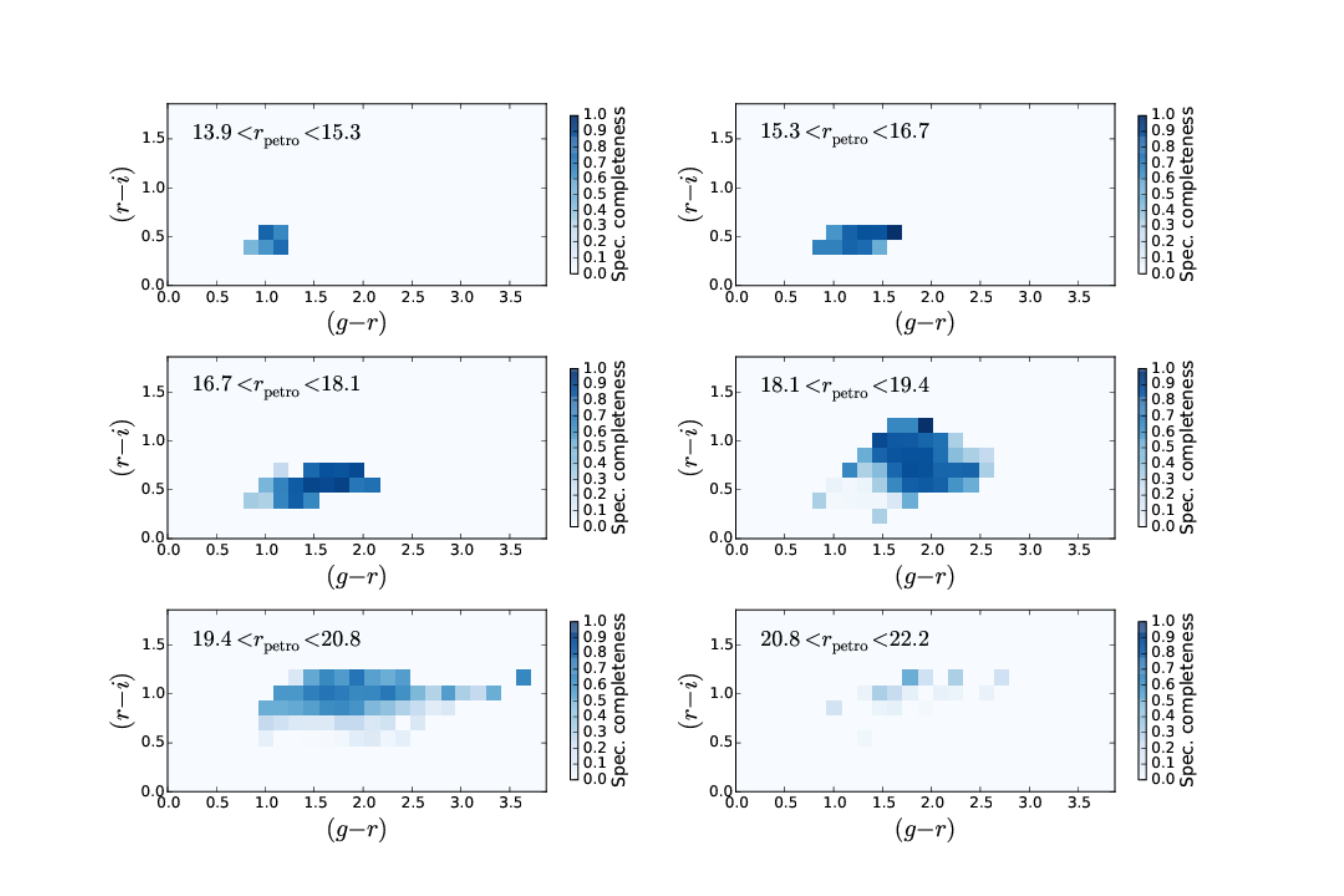}
\caption[]{The spectroscopic completeness of our pair sample as a function of $(g - r)$ and $(r - i)$ colours for six $r_{\mathrm{petro}}$ magnitude slices for the close pairs in the MGS, LRG and BOSS surveys. The colour bar to the right of each panel represents the value of the $C$-correction for each voxel. At $r_{\mathrm{petro}} \gtrsim 20.8$ mag (last panel) the spectroscopic completeness of the sample drops to $\sim0$. See text for details.}
\label{fig:C_rpetro_gr_ri_all_Ngals_MGS_LRG_BOSS}
\end{figure*}

We find that the completeness of the MGS, LRG and BOSS samples are fairly constant with $r_{\mathrm{ang}}$. The $C$-correction of the close pairs is therefore only dependent on three variables, i.e. the $r_{\mathrm{petro}}$ magnitudes, the $(g - r)$ and $(r - i)$ colours. The $C$-correction of the pairs are determined by constructing a three-dimensional cube where each of the three variables represent an axis (see Fig. \ref{fig:C_rpetro_gr_ri_all_Ngals_MGS_LRG_BOSS}). The galaxies in our close pair sample are divided into $r_{\mathrm{petro}}$ magnitude slices of $\sim1.4$ mag to ensure a smooth $C$-correction transition between the voxels. The $(g - r)$ vs. $(r - i)$ colour diagram of the galaxies in each magnitude slice is then plotted. Thereafter, Eq. \ref{eq:spec_comp_fac} is used to determine the $C$-correction of each voxel.

The spectroscopic completeness of the pair sample is expected to drop around $m_{\mathrm{i}} = 19.9$ mag since this is the magnitude limit of the BOSS survey (see Table~\ref{table:spec_sample_selec}). This roughly corresponds to $m_{\mathrm{r}} = 20.8$ mag (refer to last panel of Fig. \ref{fig:C_rpetro_gr_ri_all_Ngals_MGS_LRG_BOSS}). For any galaxy with a magnitude fainter than this, we set the $C$-correction equal to zero. As previously mentioned spectroscopy is vital in order to determine which companions are bound to their host BCGs. To ensure that we have a sample with a high spectroscopic completeness with which the BCG pair fraction can be determined, we further restrict our close pair sample to only include galaxies brighter than the BOSS magnitude limit. At this magnitude, we expect to be complete for major mergers (see Fig. \ref{fig:mag_lim_21_5}). The pair fraction is then determined using Eq. \ref{eq:merg_frac_c_corr}.

\bibliographystyle{mnras}
\bibliography{ms}  

\begin{thebibliography}{}
\makeatletter
\relax
\def\mn@urlcharsother{\let\do\@makeother \do\$\do\&\do\#\do\^\do\_\do\%\do\~}
\def\mn@doi{\begingroup\mn@urlcharsother \@ifnextchar [ {\mn@doi@}
  {\mn@doi@[]}}
\def\mn@doi@[#1]#2{\def\@tempa{#1}\ifx\@tempa\@empty \href
  {http://dx.doi.org/#2} {doi:#2}\else \href {http://dx.doi.org/#2} {#1}\fi
  \endgroup}
\def\mn@eprint#1#2{\mn@eprint@#1:#2::\@nil}
\def\mn@eprint@arXiv#1{\href {http://arxiv.org/abs/#1} {{\tt arXiv:#1}}}
\def\mn@eprint@dblp#1{\href {http://dblp.uni-trier.de/rec/bibtex/#1.xml}
  {dblp:#1}}
\def\mn@eprint@#1:#2:#3:#4\@nil{\def\@tempa {#1}\def\@tempb {#2}\def\@tempc
  {#3}\ifx \@tempc \@empty \let \@tempc \@tempb \let \@tempb \@tempa \fi \ifx
  \@tempb \@empty \def\@tempb {arXiv}\fi \@ifundefined
  {mn@eprint@\@tempb}{\@tempb:\@tempc}{\expandafter \expandafter \csname
  mn@eprint@\@tempb\endcsname \expandafter{\@tempc}}}

\bibitem[\protect\citeauthoryear{{Abazajian} et~al.,}{{Abazajian}
  et~al.}{2004}]{Abazajian2004}
{Abazajian} K.,  et~al., 2004, \mn@doi [\aj] {10.1086/421365}, \href
  {http://adsabs.harvard.edu/abs/2004AJ....128..502A} {128, 502}

\bibitem[\protect\citeauthoryear{{Abazajian} et~al.,}{{Abazajian}
  et~al.}{2009}]{Abazajian2009}
{Abazajian} K.~N.,  et~al., 2009, \mn@doi [\apjs]
  {10.1088/0067-0049/182/2/543}, \href
  {http://adsabs.harvard.edu/abs/2009ApJS..182..543A} {182, 543}

\bibitem[\protect\citeauthoryear{{Adelman-McCarthy} et~al.,}{{Adelman-McCarthy}
  et~al.}{2007}]{Adelman-McCarthy2007}
{Adelman-McCarthy} J.~K.,  et~al., 2007, \mn@doi [\apjs] {10.1086/518864},
  \href {http://adsabs.harvard.edu/abs/2007ApJS..172..634A} {172, 634}

\bibitem[\protect\citeauthoryear{{Ahn} et~al.,}{{Ahn} et~al.}{2012}]{Ahn2012}
{Ahn} C.~P.,  et~al., 2012, \mn@doi [\apjs] {10.1088/0067-0049/203/2/21}, \href
  {http://adsabs.harvard.edu/abs/2012ApJS..203...21A} {203, 21}

\bibitem[\protect\citeauthoryear{{Aihara} et~al.,}{{Aihara}
  et~al.}{2011}]{Aihara2011}
{Aihara} H.,  et~al., 2011, \mn@doi [\apjs] {10.1088/0067-0049/193/2/29}, \href
  {http://adsabs.harvard.edu/abs/2011ApJS..193...29A} {193, 29}

\bibitem[\protect\citeauthoryear{Alam et~al.,}{Alam et~al.}{2015}]{Alam2015}
Alam S.,  et~al., 2015, \apjs, 219, 12

\bibitem[\protect\citeauthoryear{{Aragon-Salamanca}, {Baugh}  \&
  {Kauffmann}}{{Aragon-Salamanca} et~al.}{1998}]{Aragon-Salamanca1998}
{Aragon-Salamanca} A.,  {Baugh} C.~M.,   {Kauffmann} G.,  1998, \mn@doi
  [\mnras] {10.1046/j.1365-8711.1998.01495.x}, \href
  {http://adsabs.harvard.edu/abs/1998MNRAS.297..427A} {297, 427}

\bibitem[\protect\citeauthoryear{{Beers} \& {Tonry}}{{Beers} \&
  {Tonry}}{1986}]{Beers1986}
{Beers} T.~C.,  {Tonry} J.~L.,  1986, \mn@doi [\apj] {10.1086/163833}, \href
  {http://adsabs.harvard.edu/abs/1986ApJ...300..557B} {300, 557}

\bibitem[\protect\citeauthoryear{{Bell}, {Phleps}, {Somerville}, {Wolf},
  {Borch}  \& {Meisenheimer}}{{Bell} et~al.}{2006}]{Bell2006a}
{Bell} E.~F.,  {Phleps} S.,  {Somerville} R.~S.,  {Wolf} C.,  {Borch} A.,
  {Meisenheimer} K.,  2006, \mn@doi [\apj] {10.1086/508408}, \href
  {http://adsabs.harvard.edu/abs/2006ApJ...652..270B} {652, 270}

\bibitem[\protect\citeauthoryear{{Bernardi}, {Hyde}, {Sheth}, {Miller}  \&
  {Nichol}}{{Bernardi} et~al.}{2007}]{Bernardi2007}
{Bernardi} M.,  {Hyde} J.~B.,  {Sheth} R.~K.,  {Miller} C.~J.,   {Nichol}
  R.~C.,  2007, \mn@doi [\aj] {10.1086/511783}, \href
  {http://adsabs.harvard.edu/abs/2007AJ....133.1741B} {133, 1741}

\bibitem[\protect\citeauthoryear{{Bertone} \& {Conselice}}{{Bertone} \&
  {Conselice}}{2009}]{Bertone2009}
{Bertone} S.,  {Conselice} C.~J.,  2009, \mn@doi [\mnras]
  {10.1111/j.1365-2966.2009.14916.x}, \href
  {http://adsabs.harvard.edu/abs/2009MNRAS.396.2345B} {396, 2345}

\bibitem[\protect\citeauthoryear{{Blanton} \& {Roweis}}{{Blanton} \&
  {Roweis}}{2007}]{Blanton2007}
{Blanton} M.~R.,  {Roweis} S.,  2007, \mn@doi [\aj] {10.1086/510127}, \href
  {http://adsabs.harvard.edu/abs/2007AJ....133..734B} {133, 734}

\bibitem[\protect\citeauthoryear{{Blanton}, {Lin}, {Lupton}, {Maley}, {Young},
  {Zehavi}  \& {Loveday}}{{Blanton} et~al.}{2003}]{Blanton2003}
{Blanton} M.~R.,  {Lin} H.,  {Lupton} R.~H.,  {Maley} F.~M.,  {Young} N.,
  {Zehavi} I.,   {Loveday} J.,  2003, \mn@doi [\aj] {10.1086/344761}, \href
  {http://adsabs.harvard.edu/abs/2003AJ....125.2276B} {125, 2276}

\bibitem[\protect\citeauthoryear{{Brough}, {Couch}, {Collins}, {Jarrett},
  {Burke}  \& {Mann}}{{Brough} et~al.}{2008}]{Brough2008}
{Brough} S.,  {Couch} W.~J.,  {Collins} C.~A.,  {Jarrett} T.,  {Burke} D.~J.,
  {Mann} R.~G.,  2008, \mn@doi [\mnras] {10.1111/j.1745-3933.2008.00442.x},
  \href {http://adsabs.harvard.edu/abs/2008MNRAS.385L.103B} {385, L103}

\bibitem[\protect\citeauthoryear{{Brough}, {Tran}, {Sharp}, {von der Linden}
  \& {Couch}}{{Brough} et~al.}{2011}]{Brough2011}
{Brough} S.,  {Tran} K.-V.,  {Sharp} R.~G.,  {von der Linden} A.,   {Couch}
  W.~J.,  2011, \mn@doi [\mnras] {10.1111/j.1745-3933.2011.01060.x}, \href
  {http://adsabs.harvard.edu/abs/2011MNRAS.414L..80B} {414, L80}

\bibitem[\protect\citeauthoryear{{Bruzual} \& {Charlot}}{{Bruzual} \&
  {Charlot}}{2003}]{Bruzual2003}
{Bruzual} G.,  {Charlot} S.,  2003, \mn@doi [\mnras]
  {10.1046/j.1365-8711.2003.06897.x}, \href
  {http://adsabs.harvard.edu/abs/2003MNRAS.344.1000B} {344, 1000}

\bibitem[\protect\citeauthoryear{{Buckley}, {Swart}  \& {Meiring}}{{Buckley}
  et~al.}{2006}]{Buckley2006}
{Buckley} D.~A.~H.,  {Swart} G.~P.,   {Meiring} J.~G.,  2006, in Society of
  Photo-Optical Instrumentation Engineers (SPIE) Conference Series. p.~0,
  \mn@doi{10.1117/12.673750}

\bibitem[\protect\citeauthoryear{{Bundy}, {Fukugita}, {Ellis}, {Targett},
  {Belli}  \& {Kodama}}{{Bundy} et~al.}{2009}]{Bundy2009}
{Bundy} K.,  {Fukugita} M.,  {Ellis} R.~S.,  {Targett} T.~A.,  {Belli} S.,
  {Kodama} T.,  2009, \mn@doi [\apj] {10.1088/0004-637X/697/2/1369}, \href
  {http://adsabs.harvard.edu/abs/2009ApJ...697.1369B} {697, 1369}

\bibitem[\protect\citeauthoryear{{Burbidge}}{{Burbidge}}{1975}]{Burbidge1975}
{Burbidge} G.,  1975, \mn@doi [\apjl] {10.1086/181731}, \href
  {http://adsabs.harvard.edu/abs/1975ApJ...196L...7B} {196, L7}

\bibitem[\protect\citeauthoryear{{Burgh}, {Nordsieck}, {Kobulnicky},
  {Williams}, {O'Donoghue}, {Smith}  \& {Percival}}{{Burgh}
  et~al.}{2003}]{Burgh2003}
{Burgh} E.~B.,  {Nordsieck} K.~H.,  {Kobulnicky} H.~A.,  {Williams} T.~B.,
  {O'Donoghue} D.,  {Smith} M.~P.,   {Percival} J.~W.,  2003, in {Iye} M.,
  {Moorwood} A.~F.~M.,  eds,  Society of Photo-Optical Instrumentation
  Engineers (SPIE) Conference Series Vol. 4841, Instrument Design and
  Performance for Optical/Infrared Ground-based Telescopes. pp 1463--1471,
  \mn@doi{10.1117/12.460312}

\bibitem[\protect\citeauthoryear{{Burke} \& {Collins}}{{Burke} \&
  {Collins}}{2013}]{Burke2013}
{Burke} C.,  {Collins} C.~A.,  2013, \mn@doi [\mnras] {10.1093/mnras/stt1192},
  \href {http://adsabs.harvard.edu/abs/2013MNRAS.434.2856B} {434, 2856}

\bibitem[\protect\citeauthoryear{{Burke}, {Collins}  \& {Mann}}{{Burke}
  et~al.}{2000}]{Burke2000}
{Burke} D.~J.,  {Collins} C.~A.,   {Mann} R.~G.,  2000, \mn@doi [\apjl]
  {10.1086/312579}, \href {http://adsabs.harvard.edu/abs/2000ApJ...532L.105B}
  {532, L105}

\bibitem[\protect\citeauthoryear{{Burke}, {Collins}, {Stott}  \&
  {Hilton}}{{Burke} et~al.}{2012}]{Burke2012}
{Burke} C.,  {Collins} C.~A.,  {Stott} J.~P.,   {Hilton} M.,  2012, \mn@doi
  [\mnras] {10.1111/j.1365-2966.2012.21555.x}, \href
  {http://adsabs.harvard.edu/abs/2012MNRAS.425.2058B} {425, 2058}

\bibitem[\protect\citeauthoryear{{Burke}, {Hilton}  \& {Collins}}{{Burke}
  et~al.}{2015}]{Burke2015}
{Burke} C.,  {Hilton} M.,   {Collins} C.,  2015, \mn@doi [\mnras]
  {10.1093/mnras/stv450}, \href
  {http://adsabs.harvard.edu/abs/2015MNRAS.449.2353B} {449, 2353}

\bibitem[\protect\citeauthoryear{{Cameron}}{{Cameron}}{2011}]{Cameron2011}
{Cameron} E.,  2011, \mn@doi [\pasa] {10.1071/AS10046}, \href
  {http://adsabs.harvard.edu/abs/2011PASA...28..128C} {28, 128}

\bibitem[\protect\citeauthoryear{{Cardiel}, {Gorgas}  \&
  {Aragon-Salamanca}}{{Cardiel} et~al.}{1995}]{Cardiel1995}
{Cardiel} N.,  {Gorgas} J.,   {Aragon-Salamanca} A.,  1995, \mn@doi [\mnras]
  {10.1093/mnras/277.2.502}, \href
  {http://adsabs.harvard.edu/abs/1995MNRAS.277..502C} {277, 502}

\bibitem[\protect\citeauthoryear{{Carlberg} et~al.,}{{Carlberg}
  et~al.}{2000}]{Carlberg2000}
{Carlberg} R.~G.,  et~al., 2000, \mn@doi [\apjl] {10.1086/312560}, \href
  {http://adsabs.harvard.edu/abs/2000ApJ...532L...1C} {532, L1}

\bibitem[\protect\citeauthoryear{{Chabrier}}{{Chabrier}}{2003}]{Chabrier2003}
{Chabrier} G.,  2003, \mn@doi [\pasp] {10.1086/376392}, \href
  {http://adsabs.harvard.edu/abs/2003PASP..115..763C} {115, 763}

\bibitem[\protect\citeauthoryear{{Collins} et~al.,}{{Collins}
  et~al.}{2009}]{Collins2009}
{Collins} C.~A.,  et~al., 2009, \mn@doi [\nat] {10.1038/nature07865}, \href
  {http://adsabs.harvard.edu/abs/2009Natur.458..603C} {458, 603}

\bibitem[\protect\citeauthoryear{{Conroy}, {Ho}  \& {White}}{{Conroy}
  et~al.}{2007a}]{Conroy2007a}
{Conroy} C.,  {Ho} S.,   {White} M.,  2007a, \mn@doi [\mnras]
  {10.1111/j.1365-2966.2007.12033.x}, \href
  {http://adsabs.harvard.edu/abs/2007MNRAS.379.1491C} {379, 1491}

\bibitem[\protect\citeauthoryear{{Conroy}, {Wechsler}  \& {Kravtsov}}{{Conroy}
  et~al.}{2007b}]{Conroy2007}
{Conroy} C.,  {Wechsler} R.~H.,   {Kravtsov} A.~V.,  2007b, \mn@doi [\apj]
  {10.1086/521425}, \href {http://adsabs.harvard.edu/abs/2007ApJ...668..826C}
  {668, 826}

\bibitem[\protect\citeauthoryear{{Conselice}}{{Conselice}}{2009}]{Conselice2009a}
{Conselice} C.~J.,  2009, \mn@doi [\mnras] {10.1111/j.1745-3933.2009.00708.x},
  \href {http://adsabs.harvard.edu/abs/2009MNRAS.399L..16C} {399, L16}

\bibitem[\protect\citeauthoryear{{Contini}, {De Lucia}, {Villalobos}  \&
  {Borgani}}{{Contini} et~al.}{2014}]{Contini2014}
{Contini} E.,  {De Lucia} G.,  {Villalobos} {\'A}.,   {Borgani} S.,  2014,
  \mn@doi [\mnras] {10.1093/mnras/stt2174}, \href
  {http://adsabs.harvard.edu/abs/2014MNRAS.437.3787C} {437, 3787}

\bibitem[\protect\citeauthoryear{{Crawford} et~al.,}{{Crawford}
  et~al.}{2010}]{Crawford2010}
{Crawford} S.~M.,  et~al., 2010, in Society of Photo-Optical Instrumentation
  Engineers (SPIE) Conference Series. p.~25, \mn@doi{10.1117/12.857000}

\bibitem[\protect\citeauthoryear{{De Lucia} \& {Blaizot}}{{De Lucia} \&
  {Blaizot}}{2007}]{DeLucia2007}
{De Lucia} G.,  {Blaizot} J.,  2007, \mn@doi [\mnras]
  {10.1111/j.1365-2966.2006.11287.x}, \href
  {http://adsabs.harvard.edu/abs/2007MNRAS.375....2D} {375, 2}

\bibitem[\protect\citeauthoryear{{DeMaio}, {Gonzalez}, {Zabludoff}, {Zaritsky}
  \& {Brada{\v c}}}{{DeMaio} et~al.}{2015}]{DeMaio2015}
{DeMaio} T.,  {Gonzalez} A.~H.,  {Zabludoff} A.,  {Zaritsky} D.,   {Brada{\v
  c}} M.,  2015, \mn@doi [\mnras] {10.1093/mnras/stv033}, \href
  {http://adsabs.harvard.edu/abs/2015MNRAS.448.1162D} {448, 1162}

\bibitem[\protect\citeauthoryear{{Edge}}{{Edge}}{1991}]{Edge1991}
{Edge} A.~C.,  1991, \mnras, \href
  {http://adsabs.harvard.edu/abs/1991MNRAS.250..103E} {250, 103}

\bibitem[\protect\citeauthoryear{{Edwards} \& {Patton}}{{Edwards} \&
  {Patton}}{2012}]{Edwards2012}
{Edwards} L.~O.~V.,  {Patton} D.~R.,  2012, \mn@doi [\mnras]
  {10.1111/j.1365-2966.2012.21457.x}, \href
  {http://adsabs.harvard.edu/abs/2012MNRAS.425..287E} {425, 287}

\bibitem[\protect\citeauthoryear{{Eisenstein} et~al.,}{{Eisenstein}
  et~al.}{2001}]{Eisenstein2001}
{Eisenstein} D.~J.,  et~al., 2001, \mn@doi [\aj] {10.1086/323717}, \href
  {http://adsabs.harvard.edu/abs/2001AJ....122.2267E} {122, 2267}

\bibitem[\protect\citeauthoryear{{Ellison}, {Mendel}, {Scudder}, {Patton}  \&
  {Palmer}}{{Ellison} et~al.}{2013}]{Ellison2013}
{Ellison} S.~L.,  {Mendel} J.~T.,  {Scudder} J.~M.,  {Patton} D.~R.,   {Palmer}
  M.~J.~D.,  2013, \mn@doi [\mnras] {10.1093/mnras/sts546}, \href
  {http://adsabs.harvard.edu/abs/2013MNRAS.430.3128E} {430, 3128}

\bibitem[\protect\citeauthoryear{{Fakhouri}, {Ma}  \&
  {Boylan-Kolchin}}{{Fakhouri} et~al.}{2010}]{Fakhouri2010}
{Fakhouri} O.,  {Ma} C.-P.,   {Boylan-Kolchin} M.,  2010, \mn@doi [\mnras]
  {10.1111/j.1365-2966.2010.16859.x}, \href
  {http://adsabs.harvard.edu/abs/2010MNRAS.406.2267F} {406, 2267}

\bibitem[\protect\citeauthoryear{{Fraser-McKelvie}, {Brown}  \&
  {Pimbblet}}{{Fraser-McKelvie} et~al.}{2014}]{Fraser-McKelvie2014}
{Fraser-McKelvie} A.,  {Brown} M.~J.~I.,   {Pimbblet} K.~A.,  2014, \mn@doi
  [\mnras] {10.1093/mnrasl/slu117}, \href
  {http://adsabs.harvard.edu/abs/2014MNRAS.444L..63F} {444, L63}

\bibitem[\protect\citeauthoryear{{Gonzalez}, {Zabludoff}, {Zaritsky}  \&
  {Dalcanton}}{{Gonzalez} et~al.}{2000}]{Gonzalez2000}
{Gonzalez} A.~H.,  {Zabludoff} A.~I.,  {Zaritsky} D.,   {Dalcanton} J.~J.,
  2000, \mn@doi [\apj] {10.1086/308985}, \href
  {http://adsabs.harvard.edu/abs/2000ApJ...536..561G} {536, 561}

\bibitem[\protect\citeauthoryear{{Gonzalez}, {Zabludoff}  \&
  {Zaritsky}}{{Gonzalez} et~al.}{2003}]{Gonzalez2003}
{Gonzalez} A.~H.,  {Zabludoff} A.~I.,   {Zaritsky} D.,  2003, \mn@doi [\apss]
  {10.1023/A:1024649423503}, \href
  {http://adsabs.harvard.edu/abs/2003Ap%26SS.285...67G} {285, 67}

\bibitem[\protect\citeauthoryear{{Hansen}, {Sheldon}, {Wechsler}  \&
  {Koester}}{{Hansen} et~al.}{2009}]{Hansen2009}
{Hansen} S.~M.,  {Sheldon} E.~S.,  {Wechsler} R.~H.,   {Koester} B.~P.,  2009,
  \mn@doi [\apj] {10.1088/0004-637X/699/2/1333}, \href
  {http://adsabs.harvard.edu/abs/2009ApJ...699.1333H} {699, 1333}

\bibitem[\protect\citeauthoryear{{Hao} et~al.,}{{Hao} et~al.}{2010}]{Hao2010}
{Hao} J.,  et~al., 2010, \mn@doi [\apjs] {10.1088/0067-0049/191/2/254}, \href
  {http://adsabs.harvard.edu/abs/2010ApJS..191..254H} {191, 254}

\bibitem[\protect\citeauthoryear{{Hausman} \& {Ostriker}}{{Hausman} \&
  {Ostriker}}{1978}]{Hausman1978}
{Hausman} M.~A.,  {Ostriker} J.~P.,  1978, \mn@doi [\apj] {10.1086/156380},
  \href {http://adsabs.harvard.edu/abs/1978ApJ...224..320H} {224, 320}

\bibitem[\protect\citeauthoryear{{Hern{\'a}ndez-Toledo}, {Avila-Reese},
  {Conselice}  \& {Puerari}}{{Hern{\'a}ndez-Toledo}
  et~al.}{2005}]{Hernandez-Toledo2005}
{Hern{\'a}ndez-Toledo} H.~M.,  {Avila-Reese} V.,  {Conselice} C.~J.,
  {Puerari} I.,  2005, \mn@doi [\aj] {10.1086/427134}, \href
  {http://adsabs.harvard.edu/abs/2005AJ....129..682H} {129, 682}

\bibitem[\protect\citeauthoryear{{Hsieh}, {Yee}, {Lin}, {Gladders}  \&
  {Gilbank}}{{Hsieh} et~al.}{2008}]{Hsieh2008}
{Hsieh} B.~C.,  {Yee} H.~K.~C.,  {Lin} H.,  {Gladders} M.~D.,   {Gilbank}
  D.~G.,  2008, \mn@doi [\apj] {10.1086/589140}, \href
  {http://adsabs.harvard.edu/abs/2008ApJ...683...33H} {683, 33}

\bibitem[\protect\citeauthoryear{{Jogee} et~al.,}{{Jogee}
  et~al.}{2009}]{Jogee2009}
{Jogee} S.,  et~al., 2009, \mn@doi [\apj] {10.1088/0004-637X/697/2/1971}, \href
  {http://adsabs.harvard.edu/abs/2009ApJ...697.1971J} {697, 1971}

\bibitem[\protect\citeauthoryear{{Jones} \& {Forman}}{{Jones} \&
  {Forman}}{1984}]{Jones1984}
{Jones} C.,  {Forman} W.,  1984, \mn@doi [\apj] {10.1086/161591}, \href
  {http://adsabs.harvard.edu/abs/1984ApJ...276...38J} {276, 38}

\bibitem[\protect\citeauthoryear{{Jones} \& {Forman}}{{Jones} \&
  {Forman}}{1999}]{Jones1999}
{Jones} C.,  {Forman} W.,  1999, \mn@doi [\apj] {10.1086/306646}, \href
  {http://adsabs.harvard.edu/abs/1999ApJ...511...65J} {511, 65}

\bibitem[\protect\citeauthoryear{{Kartaltepe} et~al.,}{{Kartaltepe}
  et~al.}{2007}]{Kartaltepe2007}
{Kartaltepe} J.~S.,  et~al., 2007, \mn@doi [\apjs] {10.1086/519953}, \href
  {http://adsabs.harvard.edu/abs/2007ApJS..172..320K} {172, 320}

\bibitem[\protect\citeauthoryear{{Kauffmann}, {Li}  \& {Heckman}}{{Kauffmann}
  et~al.}{2010}]{Kauffmann2010}
{Kauffmann} G.,  {Li} C.,   {Heckman} T.~M.,  2010, \mn@doi [\mnras]
  {10.1111/j.1365-2966.2010.17337.x}, \href
  {http://adsabs.harvard.edu/abs/2010MNRAS.409..491K} {409, 491}

\bibitem[\protect\citeauthoryear{{Keenan} et~al.,}{{Keenan}
  et~al.}{2014}]{Keenan2014}
{Keenan} R.~C.,  et~al., 2014, \mn@doi [\apj] {10.1088/0004-637X/795/2/157},
  \href {http://adsabs.harvard.edu/abs/2014ApJ...795..157K} {795, 157}

\bibitem[\protect\citeauthoryear{{Kitzbichler} \& {White}}{{Kitzbichler} \&
  {White}}{2008}]{Kitzbichler2008}
{Kitzbichler} M.~G.,  {White} S.~D.~M.,  2008, \mn@doi [\mnras]
  {10.1111/j.1365-2966.2008.13873.x}, \href
  {http://adsabs.harvard.edu/abs/2008MNRAS.391.1489K} {391, 1489}

\bibitem[\protect\citeauthoryear{{Kobulnicky}, {Nordsieck}, {Burgh}, {Smith},
  {Percival}, {Williams}  \& {O'Donoghue}}{{Kobulnicky}
  et~al.}{2003}]{Kobulnicky2003}
{Kobulnicky} H.~A.,  {Nordsieck} K.~H.,  {Burgh} E.~B.,  {Smith} M.~P.,
  {Percival} J.~W.,  {Williams} T.~B.,   {O'Donoghue} D.,  2003, in {Iye} M.,
  {Moorwood} A.~F.~M.,  eds,  \procspie Vol. 4841, Instrument Design and
  Performance for Optical/Infrared Ground-based Telescopes. pp 1634--1644,
  \mn@doi{10.1117/12.460315}

\bibitem[\protect\citeauthoryear{{Krick} \& {Bernstein}}{{Krick} \&
  {Bernstein}}{2007}]{Krick2007}
{Krick} J.~E.,  {Bernstein} R.~A.,  2007, \mn@doi [\aj] {10.1086/518787}, \href
  {http://adsabs.harvard.edu/abs/2007AJ....134..466K} {134, 466}

\bibitem[\protect\citeauthoryear{{Laporte}, {White}, {Naab}  \&
  {Gao}}{{Laporte} et~al.}{2013}]{Laporte2013}
{Laporte} C.~F.~P.,  {White} S.~D.~M.,  {Naab} T.,   {Gao} L.,  2013, \mn@doi
  [\mnras] {10.1093/mnras/stt912}, \href
  {http://adsabs.harvard.edu/abs/2013MNRAS.435..901L} {435, 901}

\bibitem[\protect\citeauthoryear{{Lauer} et~al.,}{{Lauer}
  et~al.}{2007}]{Lauer2007}
{Lauer} T.~R.,  et~al., 2007, \mn@doi [\apj] {10.1086/518223}, \href
  {http://adsabs.harvard.edu/abs/2007ApJ...662..808L} {662, 808}

\bibitem[\protect\citeauthoryear{{Le F{\`e}vre} et~al.,}{{Le F{\`e}vre}
  et~al.}{2000}]{LeF`evre2000}
{Le F{\`e}vre} O.,  et~al., 2000, \mn@doi [\mnras]
  {10.1046/j.1365-8711.2000.03083.x}, \href
  {http://adsabs.harvard.edu/abs/2000MNRAS.311..565L} {311, 565}

\bibitem[\protect\citeauthoryear{{Lidman} et~al.,}{{Lidman}
  et~al.}{2012}]{Lidman2012}
{Lidman} C.,  et~al., 2012, \mn@doi [\mnras]
  {10.1111/j.1365-2966.2012.21984.x}, \href
  {http://adsabs.harvard.edu/abs/2012MNRAS.427..550L} {427, 550}

\bibitem[\protect\citeauthoryear{{Lidman} et~al.,}{{Lidman}
  et~al.}{2013}]{Lidman2013}
{Lidman} C.,  et~al., 2013, \mn@doi [\mnras] {10.1093/mnras/stt777}, \href
  {http://adsabs.harvard.edu/abs/2013MNRAS.433..825L} {433, 825}

\bibitem[\protect\citeauthoryear{{Lin} et~al.,}{{Lin} et~al.}{2004}]{Lin2004a}
{Lin} L.,  et~al., 2004, \mn@doi [\apjl] {10.1086/427183}, \href
  {http://adsabs.harvard.edu/abs/2004ApJ...617L...9L} {617, L9}

\bibitem[\protect\citeauthoryear{{Lin} et~al.,}{{Lin} et~al.}{2008}]{Lin2008}
{Lin} L.,  et~al., 2008, \mn@doi [\apj] {10.1086/587928}, \href
  {http://adsabs.harvard.edu/abs/2008ApJ...681..232L} {681, 232}

\bibitem[\protect\citeauthoryear{{Lin}, {Brodwin}, {Gonzalez}, {Bode},
  {Eisenhardt}, {Stanford}  \& {Vikhlinin}}{{Lin} et~al.}{2013}]{Lin2013}
{Lin} Y.-T.,  {Brodwin} M.,  {Gonzalez} A.~H.,  {Bode} P.,  {Eisenhardt}
  P.~R.~M.,  {Stanford} S.~A.,   {Vikhlinin} A.,  2013, \mn@doi [\apj]
  {10.1088/0004-637X/771/1/61}, \href
  {http://adsabs.harvard.edu/abs/2013ApJ...771...61L} {771, 61}

\bibitem[\protect\citeauthoryear{{Liu}, {Mao}, {Deng}, {Xia}  \& {Wen}}{{Liu}
  et~al.}{2009}]{Liu2009}
{Liu} F.~S.,  {Mao} S.,  {Deng} Z.~G.,  {Xia} X.~Y.,   {Wen} Z.~L.,  2009,
  \mn@doi [\mnras] {10.1111/j.1365-2966.2009.14907.x}, \href
  {http://adsabs.harvard.edu/abs/2009MNRAS.396.2003L} {396, 2003}

\bibitem[\protect\citeauthoryear{{Liu}, {Mao}  \& {Meng}}{{Liu}
  et~al.}{2012}]{Liu2012}
{Liu} F.~S.,  {Mao} S.,   {Meng} X.~M.,  2012, \mn@doi [\mnras]
  {10.1111/j.1365-2966.2012.20886.x}, \href
  {http://adsabs.harvard.edu/abs/2012MNRAS.423..422L} {423, 422}

\bibitem[\protect\citeauthoryear{{Liu}, {Lei}, {Meng}  \& {Jiang}}{{Liu}
  et~al.}{2015}]{Liu2015}
{Liu} F.~S.,  {Lei} F.~J.,  {Meng} X.~M.,   {Jiang} D.~F.,  2015, \mn@doi
  [\mnras] {10.1093/mnras/stu2543}, \href
  {http://adsabs.harvard.edu/abs/2015MNRAS.447.1491L} {447, 1491}

\bibitem[\protect\citeauthoryear{{L{\'o}pez-Sanjuan}
  et~al.,}{{L{\'o}pez-Sanjuan} et~al.}{2012}]{Lopez-Sanjuan2012}
{L{\'o}pez-Sanjuan} C.,  et~al., 2012, \mn@doi [\aap]
  {10.1051/0004-6361/201219085}, \href
  {http://adsabs.harvard.edu/abs/2012A%26A...548A...7L} {548, A7}

\bibitem[\protect\citeauthoryear{{Lotz}, {Jonsson}, {Cox}, {Croton}, {Primack},
  {Somerville}  \& {Stewart}}{{Lotz} et~al.}{2011}]{Lotz2011}
{Lotz} J.~M.,  {Jonsson} P.,  {Cox} T.~J.,  {Croton} D.,  {Primack} J.~R.,
  {Somerville} R.~S.,   {Stewart} K.,  2011, \mn@doi [\apj]
  {10.1088/0004-637X/742/2/103}, \href
  {http://adsabs.harvard.edu/abs/2011ApJ...742..103L} {742, 103}

\bibitem[\protect\citeauthoryear{{Loubser}, {Babul}, {Hoekstra}, {Mahdavi},
  {Donahue}, {Bildfell}  \& {Voit}}{{Loubser} et~al.}{2016}]{Loubser2016}
{Loubser} S.~I.,  {Babul} A.,  {Hoekstra} H.,  {Mahdavi} A.,  {Donahue} M.,
  {Bildfell} C.,   {Voit} G.~M.,  2016, \mn@doi [\mnras]
  {10.1093/mnras/stv2784}, \href
  {http://adsabs.harvard.edu/abs/2016MNRAS.456.1565L} {456, 1565}

\bibitem[\protect\citeauthoryear{{Lu}, {Gilbank}, {Balogh}  \& {Bognat}}{{Lu}
  et~al.}{2009}]{Lu2009}
{Lu} T.,  {Gilbank} D.~G.,  {Balogh} M.~L.,   {Bognat} A.,  2009, \mn@doi
  [\mnras] {10.1111/j.1365-2966.2009.15418.x}, \href
  {http://adsabs.harvard.edu/abs/2009MNRAS.399.1858L} {399, 1858}

\bibitem[\protect\citeauthoryear{{Matthews}, {Morgan}  \& {Schmidt}}{{Matthews}
  et~al.}{1964}]{Matthews1964}
{Matthews} T.~A.,  {Morgan} W.~W.,   {Schmidt} M.,  1964, \mn@doi [\apj]
  {10.1086/147890}, \href {http://adsabs.harvard.edu/abs/1964ApJ...140...35M}
  {140, 35}

\bibitem[\protect\citeauthoryear{{McIntosh}, {Guo}, {Hertzberg}, {Katz}, {Mo},
  {van den Bosch}  \& {Yang}}{{McIntosh} et~al.}{2008}]{McIntosh2008}
{McIntosh} D.~H.,  {Guo} Y.,  {Hertzberg} J.,  {Katz} N.,  {Mo} H.~J.,  {van
  den Bosch} F.~C.,   {Yang} X.,  2008, \mn@doi [\mnras]
  {10.1111/j.1365-2966.2008.13531.x}, \href
  {http://adsabs.harvard.edu/abs/2008MNRAS.388.1537M} {388, 1537}

\bibitem[\protect\citeauthoryear{{Mihos}, {Harding}, {Feldmeier}  \&
  {Morrison}}{{Mihos} et~al.}{2005}]{Mihos2005}
{Mihos} J.~C.,  {Harding} P.,  {Feldmeier} J.,   {Morrison} H.,  2005, \mn@doi
  [\apjl] {10.1086/497030}, \href
  {http://adsabs.harvard.edu/abs/2005ApJ...631L..41M} {631, L41}

\bibitem[\protect\citeauthoryear{{Murante}, {Giovalli}, {Gerhard}, {Arnaboldi},
  {Borgani}  \& {Dolag}}{{Murante} et~al.}{2007}]{Murante2007}
{Murante} G.,  {Giovalli} M.,  {Gerhard} O.,  {Arnaboldi} M.,  {Borgani} S.,
  {Dolag} K.,  2007, \mn@doi [\mnras] {10.1111/j.1365-2966.2007.11568.x}, \href
  {http://adsabs.harvard.edu/abs/2007MNRAS.377....2M} {377, 2}

\bibitem[\protect\citeauthoryear{{Naab}, {Johansson}  \& {Ostriker}}{{Naab}
  et~al.}{2009}]{Naab2009}
{Naab} T.,  {Johansson} P.~H.,   {Ostriker} J.~P.,  2009, \mn@doi [\apjl]
  {10.1088/0004-637X/699/2/L178}, \href
  {http://adsabs.harvard.edu/abs/2009ApJ...699L.178N} {699, L178}

\bibitem[\protect\citeauthoryear{{O'Donoghue} et~al.,}{{O'Donoghue}
  et~al.}{2006}]{ODonoghue2006}
{O'Donoghue} D.,  et~al., 2006, \mn@doi [\mnras]
  {10.1111/j.1365-2966.2006.10834.x}, \href
  {http://adsabs.harvard.edu/abs/2006MNRAS.372..151O} {372, 151}

\bibitem[\protect\citeauthoryear{{Oliva-Altamirano} et~al.,}{{Oliva-Altamirano}
  et~al.}{2014}]{Oliva-Altamirano2014}
{Oliva-Altamirano} P.,  et~al., 2014, \mn@doi [\mnras] {10.1093/mnras/stu277},
  \href {http://adsabs.harvard.edu/abs/2014MNRAS.440..762O} {440, 762}

\bibitem[\protect\citeauthoryear{{Ostriker} \& {Hausman}}{{Ostriker} \&
  {Hausman}}{1977}]{Ostriker1977}
{Ostriker} J.~P.,  {Hausman} M.~A.,  1977, \mn@doi [\apjl] {10.1086/182554},
  \href {http://adsabs.harvard.edu/abs/1977ApJ...217L.125O} {217, L125}

\bibitem[\protect\citeauthoryear{{Patton}, {Carlberg}, {Marzke}, {Pritchet},
  {da Costa}  \& {Pellegrini}}{{Patton} et~al.}{2000}]{Patton2000}
{Patton} D.~R.,  {Carlberg} R.~G.,  {Marzke} R.~O.,  {Pritchet} C.~J.,  {da
  Costa} L.~N.,   {Pellegrini} P.~S.,  2000, \mn@doi [\apj] {10.1086/308907},
  \href {http://adsabs.harvard.edu/abs/2000ApJ...536..153P} {536, 153}

\bibitem[\protect\citeauthoryear{{Patton} et~al.,}{{Patton}
  et~al.}{2002}]{Patton2002}
{Patton} D.~R.,  et~al., 2002, \mn@doi [\apj] {10.1086/324543}, \href
  {http://adsabs.harvard.edu/abs/2002ApJ...565..208P} {565, 208}

\bibitem[\protect\citeauthoryear{{Pipino}, {Kaviraj}, {Bildfell}, {Babul},
  {Hoekstra}  \& {Silk}}{{Pipino} et~al.}{2009}]{Pipino2009}
{Pipino} A.,  {Kaviraj} S.,  {Bildfell} C.,  {Babul} A.,  {Hoekstra} H.,
  {Silk} J.,  2009, \mn@doi [\mnras] {10.1111/j.1365-2966.2009.14534.x}, \href
  {http://adsabs.harvard.edu/abs/2009MNRAS.395..462P} {395, 462}

\bibitem[\protect\citeauthoryear{{Puchwein}, {Springel}, {Sijacki}  \&
  {Dolag}}{{Puchwein} et~al.}{2010}]{Puchwein2010}
{Puchwein} E.,  {Springel} V.,  {Sijacki} D.,   {Dolag} K.,  2010, \mn@doi
  [\mnras] {10.1111/j.1365-2966.2010.16786.x}, \href
  {http://adsabs.harvard.edu/abs/2010MNRAS.406..936P} {406, 936}

\bibitem[\protect\citeauthoryear{{Rasmussen}, {Mulchaey}, {Bai}, {Ponman},
  {Raychaudhury}  \& {Dariush}}{{Rasmussen} et~al.}{2010}]{Rasmussen2010}
{Rasmussen} J.,  {Mulchaey} J.~S.,  {Bai} L.,  {Ponman} T.~J.,  {Raychaudhury}
  S.,   {Dariush} A.,  2010, \mn@doi [\apj] {10.1088/0004-637X/717/2/958},
  \href {http://adsabs.harvard.edu/abs/2010ApJ...717..958R} {717, 958}

\bibitem[\protect\citeauthoryear{{Rhee} \& {Latour}}{{Rhee} \&
  {Latour}}{1991}]{Rhee1991}
{Rhee} G.~F.~R.~N.,  {Latour} H.~J.,  1991, \aap, \href
  {http://adsabs.harvard.edu/abs/1991A%26A...243...38R} {243, 38}

\bibitem[\protect\citeauthoryear{{Richstone} \& {Malumuth}}{{Richstone} \&
  {Malumuth}}{1983}]{Richstone1983}
{Richstone} D.~O.,  {Malumuth} E.~M.,  1983, \mn@doi [\apj] {10.1086/160926},
  \href {http://adsabs.harvard.edu/abs/1983ApJ...268...30R} {268, 30}

\bibitem[\protect\citeauthoryear{{Robaina}, {Bell}, {van der Wel},
  {Somerville}, {Skelton}, {McIntosh}, {Meisenheimer}  \& {Wolf}}{{Robaina}
  et~al.}{2010}]{Robaina2010}
{Robaina} A.~R.,  {Bell} E.~F.,  {van der Wel} A.,  {Somerville} R.~S.,
  {Skelton} R.~E.,  {McIntosh} D.~H.,  {Meisenheimer} K.,   {Wolf} C.,  2010,
  \mn@doi [\apj] {10.1088/0004-637X/719/1/844}, \href
  {http://adsabs.harvard.edu/abs/2010ApJ...719..844R} {719, 844}

\bibitem[\protect\citeauthoryear{{Robotham} et~al.,}{{Robotham}
  et~al.}{2014}]{Robotham2014}
{Robotham} A.~S.~G.,  et~al., 2014, \mn@doi [\mnras] {10.1093/mnras/stu1604},
  \href {http://adsabs.harvard.edu/abs/2014MNRAS.444.3986R} {444, 3986}

\bibitem[\protect\citeauthoryear{{Rudick}, {Mihos}  \& {McBride}}{{Rudick}
  et~al.}{2011}]{Rudick2011}
{Rudick} C.~S.,  {Mihos} J.~C.,   {McBride} C.~K.,  2011, \mn@doi [\apj]
  {10.1088/0004-637X/732/1/48}, \href
  {http://adsabs.harvard.edu/abs/2011ApJ...732...48R} {732, 48}

\bibitem[\protect\citeauthoryear{{Rykoff} et~al.,}{{Rykoff}
  et~al.}{2012}]{Rykoff2012}
{Rykoff} E.~S.,  et~al., 2012, \mn@doi [\apj] {10.1088/0004-637X/746/2/178},
  \href {http://adsabs.harvard.edu/abs/2012ApJ...746..178R} {746, 178}

\bibitem[\protect\citeauthoryear{{Rykoff} et~al.,}{{Rykoff}
  et~al.}{2014}]{Rykoff2014}
{Rykoff} E.~S.,  et~al., 2014, \mn@doi [\apj] {10.1088/0004-637X/785/2/104},
  \href {http://adsabs.harvard.edu/abs/2014ApJ...785..104R} {785, 104}

\bibitem[\protect\citeauthoryear{{Sarazin} \& {O'Connell}}{{Sarazin} \&
  {O'Connell}}{1983}]{Sarazin1983}
{Sarazin} C.~L.,  {O'Connell} R.~W.,  1983, \mn@doi [\apj] {10.1086/160978},
  \href {http://adsabs.harvard.edu/abs/1983ApJ...268..552S} {268, 552}

\bibitem[\protect\citeauthoryear{{Schlegel}, {Finkbeiner}  \&
  {Davis}}{{Schlegel} et~al.}{1998}]{Schlegel1998}
{Schlegel} D.~J.,  {Finkbeiner} D.~P.,   {Davis} M.,  1998, \mn@doi [\apj]
  {10.1086/305772}, \href {http://adsabs.harvard.edu/abs/1998ApJ...500..525S}
  {500, 525}

\bibitem[\protect\citeauthoryear{{Schombert}}{{Schombert}}{1988}]{Schombert1988}
{Schombert} J.~M.,  1988, \mn@doi [\apj] {10.1086/166306}, \href
  {http://adsabs.harvard.edu/abs/1988ApJ...328..475S} {328, 475}

\bibitem[\protect\citeauthoryear{{Springel} et~al.,}{{Springel}
  et~al.}{2005}]{Springel2005}
{Springel} V.,  et~al., 2005, \mn@doi [\nat] {10.1038/nature03597}, \href
  {http://adsabs.harvard.edu/abs/2005Natur.435..629S} {435, 629}

\bibitem[\protect\citeauthoryear{{Stott}, {Edge}, {Smith}, {Swinbank}  \&
  {Ebeling}}{{Stott} et~al.}{2008}]{Stott2008}
{Stott} J.~P.,  {Edge} A.~C.,  {Smith} G.~P.,  {Swinbank} A.~M.,   {Ebeling}
  H.,  2008, \mn@doi [\mnras] {10.1111/j.1365-2966.2007.12807.x}, \href
  {http://adsabs.harvard.edu/abs/2008MNRAS.384.1502S} {384, 1502}

\bibitem[\protect\citeauthoryear{{Stott} et~al.,}{{Stott}
  et~al.}{2010}]{Stott2010}
{Stott} J.~P.,  et~al., 2010, \mn@doi [\apj] {10.1088/0004-637X/718/1/23},
  \href {http://adsabs.harvard.edu/abs/2010ApJ...718...23S} {718, 23}

\bibitem[\protect\citeauthoryear{{Stott} et~al.,}{{Stott}
  et~al.}{2012}]{Stott2012}
{Stott} J.~P.,  et~al., 2012, \mn@doi [\mnras]
  {10.1111/j.1365-2966.2012.20764.x}, \href
  {http://adsabs.harvard.edu/abs/2012MNRAS.422.2213S} {422, 2213}

\bibitem[\protect\citeauthoryear{{Strauss} et~al.,}{{Strauss}
  et~al.}{2002}]{Strauss2002}
{Strauss} M.~A.,  et~al., 2002, \mn@doi [\aj] {10.1086/342343}, \href
  {http://adsabs.harvard.edu/abs/2002AJ....124.1810S} {124, 1810}

\bibitem[\protect\citeauthoryear{{Tody}}{{Tody}}{1986}]{Tody1986}
{Tody} D.,  1986, in {Crawford} D.~L.,  ed.,  \procspie Vol. 627,
  Instrumentation in astronomy VI. p.~733

\bibitem[\protect\citeauthoryear{{Tody}}{{Tody}}{1993}]{Tody1993}
{Tody} D.,  1993, in {Hanisch} R.~J.,  {Brissenden} R.~J.~V.,   {Barnes} J.,
  eds,  Astronomical Society of the Pacific Conference Series Vol. 52,
  Astronomical Data Analysis Software and Systems II. p.~173

\bibitem[\protect\citeauthoryear{{Tonry}}{{Tonry}}{1987}]{Tonry1987}
{Tonry} J.~L.,  1987, in {de Zeeuw} P.~T.,  ed.,  IAU Symposium Vol. 127,
  Structure and Dynamics of Elliptical Galaxies. pp 89--96

\bibitem[\protect\citeauthoryear{{Whiley} et~al.,}{{Whiley}
  et~al.}{2008}]{Whiley2008}
{Whiley} I.~M.,  et~al., 2008, \mn@doi [\mnras]
  {10.1111/j.1365-2966.2008.13324.x}, \href
  {http://adsabs.harvard.edu/abs/2008MNRAS.387.1253W} {387, 1253}

\bibitem[\protect\citeauthoryear{{York} et~al.,}{{York}
  et~al.}{2000}]{York2000}
{York} D.~G.,  et~al., 2000, \mn@doi [\aj] {10.1086/301513}, \href
  {http://adsabs.harvard.edu/abs/2000AJ....120.1579Y} {120, 1579}

\bibitem[\protect\citeauthoryear{{Zehavi} et~al.,}{{Zehavi}
  et~al.}{2002}]{Zehavi2002}
{Zehavi} I.,  et~al., 2002, \mn@doi [\apj] {10.1086/339893}, \href
  {http://adsabs.harvard.edu/abs/2002ApJ...571..172Z} {571, 172}

\bibitem[\protect\citeauthoryear{{Zhang} et~al.,}{{Zhang}
  et~al.}{2016}]{Zhang2016}
{Zhang} Y.,  et~al., 2016, \mn@doi [\apj] {10.3847/0004-637X/816/2/98}, \href
  {http://adsabs.harvard.edu/abs/2016ApJ...816...98Z} {816, 98}

\bibitem[\protect\citeauthoryear{{de Ravel} et~al.,}{{de Ravel}
  et~al.}{2009}]{2009}
{de Ravel} L.,  et~al., 2009, \mn@doi [\aap] {10.1051/0004-6361/200810569},
  \href {http://adsabs.harvard.edu/abs/2009A%26A...498..379D} {498, 379}

\bibitem[\protect\citeauthoryear{{van Dokkum}}{{van Dokkum}}{2001}]{2001}
{van Dokkum} P.~G.,  2001, \mn@doi [\pasp] {10.1086/323894}, \href
  {http://adsabs.harvard.edu/abs/2001PASP..113.1420V} {113, 1420}

\bibitem[\protect\citeauthoryear{{von der Linden}, {Best}, {Kauffmann}  \&
  {White}}{{von der Linden} et~al.}{2007}]{2007}
{von der Linden} A.,  {Best} P.~N.,  {Kauffmann} G.,   {White} S.~D.~M.,  2007,
  \mn@doi [\mnras] {10.1111/j.1365-2966.2007.11940.x}, \href
  {http://adsabs.harvard.edu/abs/2007MNRAS.379..867V} {379, 867}

\makeatother
\end{thebibliography}

\bsp	
\label{lastpage}
\end{document}